\documentclass[lettersize, journal]{IEEEtran}

\usepackage{amsmath, amsfonts}
\usepackage{amsthm}
\usepackage{algorithmic}
\usepackage{algorithm}
\usepackage{array}
\usepackage[caption = false, font = footnotesize]{subfig}
\usepackage{textcomp}
\usepackage{stfloats}
\usepackage{url}
\usepackage{verbatim}
\usepackage{graphicx}
\usepackage{cite}
\usepackage{hyperref}
\newtheorem{pro}{Proposition}
\graphicspath{{img/}}
\usepackage{booktabs}
\usepackage{tabularx}
\usepackage{color}

\begin{document}

\title{Data Monetization Pathways and Complex Dynamic Game Equilibrium Analysis in the Energy Industry}

\author{Zongxian Wang, Jie Song
\thanks{This work was supported in part by the National Science Foundation of China (No. 72241420) and Young Scientists Fund of the National Natural Science Foundation of China (No. 72301007).}
\thanks{Zongxian Wang and Jie Song are with the Department of Industrial Engineering and Management, College of Engineering, Peking University, Beijing 100871, China (e-mail: zongxian.wang@pku.edu.cn; jie.song@pku.edu.cn).}
\thanks{Manuscript received xx xx, xxxx; revised xx xx, xxxx.}}

\markboth{Journal of \LaTeX\ Class Files, ~Vol.~14, No.~8, August~2023}%
{Shell \MakeLowercase{\textit{WANG et al.}}: Data Monetization Pathways and Game Equilibrium Analysis in the Energy Industry}


\maketitle

\begin{abstract}

As the most critical production factor in the era of the digital economy, data will have a significant impact on social production and development. Energy enterprises possess data that is interconnected with multiple industries, characterized by diverse needs, sensitivity, and long-term nature. The path to monetizing energy enterprises' data is challenging yet crucial.
This paper explores the game-theoretic aspects of the data monetization process in energy enterprises by considering the relationships between enterprises and trading platforms. We construct a class of game decision models and study their equilibrium strategies. Our analysis shows that enterprises and platforms can adjust respective benefits by regulating the wholesale price of data and the intensity of data value mining to form a benign equilibrium state. Furthermore, by integrating nonlinear dynamical theory, we discuss the dynamic characteristics present in multi-period repeated game processes. We find that decision-makers should keep the adjustment parameters and initial states within reasonable ranges in multi-period dynamic decision-making to avoid market failure. Finally, based on the theoretical and numerical analysis, we provide decision insights and recommendations for enterprise decision-making to facilitate data monetization through strategic interactions with trading platforms.

\end{abstract}


\section{Introduction}\label{sec01}
\vspace{5mm}

\IEEEPARstart{W}{ith} the rapid advancement of technology and the advent of the digital age, data has become a crucial asset in today's society\cite{Klos2023,Rocha2023}. Data is not just the carrier of information, but also the core tool for profound insights into the world and guiding decision-making. As key players in the energy industry, energy enterprises possess massive heterogeneous data resources from a variety of sources. This data encompasses the entire spectrum of the industry including production, transmission, distribution, and consumption, with extensive time range, geographical distribution, and complexity. As intelligentization and digitalization levels rise, the speed of enterprise data generation and accumulation grows exponentially, leading to increasingly exorbitant storage and processing costs. Concurrently, governments and enterprises are also gradually recognizing the value of data, managing it as a strategic asset. Therefore, modern energy firms must acknowledge the significance of data assets, instituting standardized management systems, developing analytical and mining models, and profoundly tapping the value of data to support refined operations and integrated decision-making. Data monetization can optimize the entire production and operation chain, and propel the industry toward intelligentization and green low-carbon development. Energy enterprises should seize the strategic opportunity of data monetization, accelerate digital transformation, and harness new competitive edges driven by data.

Currently, many energy enterprises have already embarked on data monetization practices and gained some benefits from it\cite{Zhang2023}. For example, as a globally renowned energy company, BP has realized the monetization of data value by providing energy industry-related data services, including energy trend analysis, market reports and energy data platforms. Shell, as an integrated energy company, can also leverage its data resources in areas like oil, natural gas and new energy to provide customized data solutions for clients. In China, the road towards enterprise data monetization has also become a hot research topic. To further promote enterprise data monetization and improve data regulation, the Chinese government is continuously pushing forward the construction of data trading platforms. Based on the practices of enterprise data monetization globally, this paper takes the perspective of energy enterprises participating in data trading, summarizes a trading model that considers both the participation of energy enterprises and third-party trading platforms, and then applies game theory to explore the relationship between enterprises and platforms, analyzing the willingness and conditions for both parties to engage in data trading. This research provides theoretical basis and practical references for energy enterprises to participate in data monetization and carry out data trading.

\begin{figure}[!htb]
	\centering
	\includegraphics[width = \linewidth]{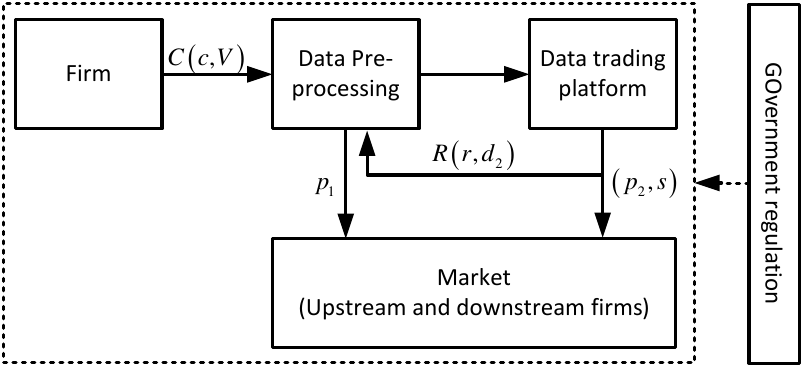}
	\caption{Model framework and decision sequence.}
	\label{fig: figframework}
\end{figure}

Since enterprise data monetization is still in the exploratory and developmental stages, existing literature mainly discusses aspects like data ownership and data security. Different from these studies, this paper attempts to study the motivations and influences of enterprise participation in data trading from the perspective of data trading models and market behaviors. We construct a game model focusing on the prices and benefits of enterprises in data trading and the role of trading platforms, without involving complex issues like data ownership and security. This helps us dissect the essence of enterprise data monetization more clearly from an economics standpoint. The framework of the model constructed in this paper is shown in Figure \ref{fig: figframework}. 

Compared with existing research, the main contributions of this paper are as follows: 
\begin{enumerate}
	\item This paper analyzes the realization path of data monetization from the perspective of enterprise participation in data trading, and reveals the intrinsic motivations and extrinsic influencing factors for enterprises to participate in data trading. The research results provide management insights for energy enterprises to better realize data monetization.
	\item The research shows that the market demand for raw data and processed data is linked to data value. Enterprises can change profit levels by adjusting inputs. Meanwhile, the government can increase the earnings of all parties in data trading by promoting the progress of data analysis technologies.
	\item This paper constructs a dynamic repeat data trading game model and analyzes the local and global stability of the system, pointing out the existence of period-doubling, coexisting attractors and other dynamical characteristics. The corresponding conclusions can provide more reasonable solutions for corporate and government decision-making to ensure stable market trading behaviors.

\end{enumerate}

The remainder of this paper proceeds as follows. Section \ref{sec02} reviews existing literature encompassing data monetization pathways, pricing strategies, and dynamic system analyses. Section \ref{sec03} delineates our baseline game-theoretic model formulation to characterize the single-stage static scenario. Section \ref{sec04} further constrcuts an enriched dynamic gaming framework to elucidate the emergent complex behavioral phenomena and properties. Finally, Section \ref{sec05} concludes by summarizing key findings and contributions from this research.

\section{Literature Review}\label{sec02}

Our research is mainly related to three streams of studies, i.e., data monetization pathways, optimal pricing strategies, and analyzing nonlinear dynamics in game scenarios.

\subsection{Data Monetization}

With the advent of the digital era, data has emerged as an important economic resource endowed with tremendous value. Realizing the commodity attribute of data assets through trustworthy sharing practices is crucial for unleashing its value across industrial sectors. As examined by Wang et al.\cite{Wang2023}, considerable challenges around rights, privacy and market mechanisms must be addressed to enable credible and sustainable data openness. 
Wang et al.\cite{Wang2020,Wang2023} examine the inherent uncertainties in renewable energy stemming from natural randomness, unlocking valuable statistical insights within the data. This reveals significant yet untouched opportunities for energy corporations to extract and monetize the data value via advanced analytics and sharing practices.

To augment the security and efficiency dimensions underlying data monetization procedures, Abbasi et al. \cite{Abbasi2023} proposed an industrial data trading framework leveraging blockchain architecture to transcend limitations marring conventional centralized data market ecosystems, including pervasive trust issues, single-point failure risks as well as pronounced security and privacy vulnerabilities.
To enable secure data transactions, some scholars have also conducted relevant research from the blockchain perspective, such as \cite{Song2023}.

Figueredo et al. \cite{Figueredo2022} proposes and validates through deployments an IoT data marketplace framework leveraging standardization to facilitate broader, controlled data sharing between suppliers and consumers, enabling wider innovation via interoperability and new ecosystem business models.
Moreover,  through rigorous multi-case examination of data commercialization across B2B settings, Ritala et al. \cite{Ritala2024} elucidated four archetypal categories of data-centered value propositions amenable for enterprise monetization, revealing the key data-related capabilities and challenges associated with selling and monetizing data into enhanced products, services, and business outcomes. 

Recently, Zhang et al. \cite{Zhang2023} develops a game-theoretic model to investigate how the emergence of direct data monetization impacts competition between high-value and low-value firms, finding that data brokers with strong analytics capabilities may hinder rather than facilitate market competition, though direct data monetization can improve consumer surplus and social welfare by enabling data sharing.
Complementarily, Bataineh et al. \cite{Bataineh2022} proposes a game theoretic model enhancing cooperation among data providers, cloud platforms and users to improve the overall surplus associated with monetizing data on the cloud by overcoming existing barriers to formally commodifying this increasingly valuable economic asset. Gonçalves \cite{Goncalves2021} developed and tested a renewable energy forecasting data marketplace using an adapted auction mechanism to incentivize collaboration between distributed data owners through data monetization, accommodating time-series covariates and continuous model updating.

Building on these works, some studies has focused on developing models and mechanisms to understand and improve outcomes associated with emerging practices of data commodification and monetization, with attention to market competition, cooperation, and incentive structures. This paper also aims to apply game theoretic approaches to investigate enterprise motivations and influences through the process of data association transformation.

\subsection{Pricing Strategy}

Applying game theory to study pricing strategies for competitive products in the market has become an important research area. Existing literature has considered many factors influencing prices, such as consumer \cite{Yang2018, Meng2021}, firm R\&D capabilities\cite{Chu2022}, and investment scales, which can all impact product pricing.
However, through a review of literature, it is found that current research combining optimal pricing strategies to study energy companies' data assetization remains scarce. Therefore, we can draw inspiration from some articles on game decision making. 

Insights from supply chain gaming literature can be leveraged, applying the methodologies to determine optimal pricing strategies for customer data assets during data assetization processes\cite{RastiBarzoki2021}. 
Moreover, Rezaei et al. \cite{Rezaei2022} aims to bridge the gap in understanding optimal pricing strategies for emerging data assetization practices in the energy industry by accounting for relevant game theoretic dynamics between companies and platforms. Findings stand to inform both theory and practice around strategically leveraging industrial data resources. 
Moreover, since this paper examines energy companies selling data through platforms, we can reference some studies from third-party platforms and the game theoretic intricacies therein, such as \cite{Feng2021,Zhen2022a,Liu2021}.

\subsection{Dynamic Game Analysis}

The application of game theory to study competitive behaviors between firms is often limited to one-shot games. However, introducing nonlinear dynamics can effectively characterize the dynamic repeated games between players. Decision makers can be fully rational or bounded rational\cite{Bischi2000,Bischi2002}. When considering different scenarios, complex dynamical phenomena can emerge during the gaming process as well. Such as biffurcation, coexisting attractiors and complex basins\cite{Bischi2005}.

Generally, game players adjust the long-term game strategies through the bounded rationality adjustment mechanism\cite{Bischi2000}. For instance, Yang et al. \cite{Yang2019} studied a Stackelberg duopoly game with bounded rationality strategy adjustments, analyzing the dynamic characteristics. Additionally, Cao et al. \cite{Cao2019} established a dynamic Cournot duopoly model with bounded rationality adjustments and consumer surplus. While nonlinear dynamics can generate valuable insights, bifurcations and chaos emerging in long-term behaviors \cite{Zhang2021a} may adversely impact decision-making;  thus, as Xiao et al. demonstrate, chaotic control methods like combined parameter perturbation and state feedback can prevent bifurcations and chaos, stabilizing the system dynamics\cite{Luo2003}.

In summary, nonlinear dynamics and bounded rationality considerations allow for capturing complex phenomena in repeated games between firms. Specific lines of study include dynamic models with adjustment mechanisms and assessments of resulting system behaviors. This affords more realism to competitive interaction research compared to one-shot, perfectly rational frameworks.

\section{Basic Game Model}\label{sec03}

To accurately depict the data trading behaviors of energy enterprises in the data market, this paper constructs a model consisting of a single energy enterprise and a data trading platform. The enterprise can either directly sell its own data assets to upstream and downstream enterprises, or conduct sales through a third-party platform. Unlike some existing platforms that merely act as intermediaries, the data trading platform in this paper is set to process and increase the value of acquired data, and pay rewards to the enterprise as agreed. This setting is more aligned with the essence of data monetization and is more conducive to motivating enterprises to proactively engage in data trading. It should be noted that this paper does not consider the model of platforms charging commissions, because that could reduce enterprise participation enthusiasm and falls outside the scope of this research. 
Overall, the model constructed in this paper reflects the core behavioral characteristics of energy enterprise data monetization under reasonable assumptions, and can provide a good analytical framework for related theoretical research.

Considering the sensitivity of energy enterprise data, enterprises need to conduct necessary processing when selling data. This paper denotes the processing cost as $ C (c, V) $, where $ c $ is the cost per unit of data, and $ V (d_1, d_2) = d_1 + d_2 $ is the data volume. However, since data processing usually follows fixed schemes, this cost can be seen as a constant. To simplify the analysis, this paper assumes the data processing cost is zero.
In addition, there is currently no unified measurement standard for data volume. Enterprises mainly divide data based on experience, so this paper does not specifically depict the measurement unit of data. It simply assumes the enterprise's unit data price is $ p_1 $, the platform can derive an expected business value of $ s $ per unit of data by analyzing the data, with a corresponding sale price of $ p_2 $. The platform pays the enterprise a reward $ r $ for each unit of data, where the data volume is $ d_2 $. These assumptions simplify the model establishment while retaining the key elements of enterprise data monetization.

The main parameters and decision variables that are used to build our models are summarized in Table \ref{tab01}.
\begin{table}[h]
	\centering\footnotesize
	\caption{Description of parameters and variables.}\label{tab01}
	\begin{tabularx}{\linewidth}{lX}
		\toprule
		Symbols & Description\\
		\hline
		\multicolumn{2}{l}{Model parameters} \\
		\hline
		 $ a $ & Potential market size, $ a > 0 $. \\
		 $ b $ & Cross price elasticity, which measures the responsiveness of demand for one supplier's product to a change in the price of another supplier's product, $ 0 < b \leq 1 $. \\
		 $ c $ & Pre-processing cost of each unit of data. Considering the scale effects, it is normalized to zero without loss of generality. \\
		 $ \theta $ & Consumer sensitivity to the mathematical expectation $ s $ of random business insights worth, $ 0 < \theta < 1 $. \\
		 $ r $ & Wholesale price of unit data, $ r > 0 $. \\
		 $ \eta $ & Coefficient for investment in business value $ s $ improvement, $ \eta > 0 $. \\
		\hline
		\multicolumn{2}{l}{Decision variables} \\
		\hline
		 $ p_1 $ & The price of data that the company sells directly, $ p_1 \geq 0 $. \\
		 $ p_2 $ & The price of data that the company sells through third-party data trading platforms, $ p_2 \geq 0 $. \\
		 $ s $ & Generally, the business value generated from each unit of data is a random variable. However, consumers only pay attention to the mathematical expectation of worth $ s $, $ s \geq 0 $.\\
		\hline
		\multicolumn{2}{l}{Other variables} \\
		\hline
		 $ \pi_1 $ & The profit that the company sells directly. \\
		 $ \pi_2 $ & The profit of the third-party data trading platforms. \\
		\bottomrule

	\end{tabularx}
\end{table}

Furthermore, based on the aforementioned assumptions and analysis, we construct the demand functions for the energy enterprise and the data trading platform, respectively: 
\begin{equation}\label{eq01}
	\begin{cases}
		d_1 = a - b\left ({p_1 - p_2} \right); \\
		d_2 = a - b\left ({p_2 - p_1} \right) + \theta s.
	\end{cases}
\end{equation}

Deriving from the demand functions, we formulate the profit functions for the enterprise and third-party platform. The enterprise's profit comprises direct sales revenue $ d_1 p_1 $ and revenue $ d_2 r $ from the platform. The platform's profit equals its sales revenue $ d_2 p_2 $ less wholesale data costs $ d_2 r $ paid to the enterprise and investment costs $ \frac{1}{2}\eta {s^2} $ to improve business value\cite{Zhang2021,Zhen2022}. A widely adopted quadratic function describes the investment costs, where $ \eta $ represents the investment coefficient determining nonlinear growth in costs as business value rises. The formulated profit functions are: 
\begin{equation}\label{eq02}
	\begin{cases}
		\pi_1 = d_1 p_1 + d_2 r; \\
		\pi_2 = d_2 (p_2 - r) - \frac{1}{2}\eta {s^2}.
	\end{cases}
\end{equation}

As both the enterprise and trading platform participate in market transactions, we assume they strategically interact to simultaneously determine equilibrium solutions. Then the equilibrium solutions are follows.
\begin{equation}\label{eq03}
	\begin{cases}\displaystyle
		p_1 = r + \frac{1}{2}a\left ({\frac{1}{b} + \frac{{3\eta }}{{3b\eta - 2{\theta ^2}}}} \right); \\ \displaystyle
		p_2 = r + \frac{{3a\eta }}{{3b\eta - 2{\theta ^2}}}; \\ \displaystyle
		s = \frac{{3a\theta }}{{3b\eta - 2{\theta ^2}}}.
	\end{cases}
\end{equation}

Based on the derived optimal equilibrium solutions, we can deduce the resultant profits for the energy enterprise and third-party trading platform accordingly.
\begin{equation}\label{eq04}
	\begin{cases} \displaystyle
		\pi_1 = \frac{{a\left ({br\left ({3b\eta - 2{\theta ^2}} \right) \left ({6b\eta - {\theta ^2}} \right) + a{{\left ({{\theta ^2} - 3b\eta } \right) }^2}} \right) }}{{b{{\left ({3b\eta - 2{\theta ^2}} \right) }^2}}}; \\ \displaystyle
		\pi_2 = \frac{{9{a^2}\eta \left ({2b\eta - {\theta ^2}} \right) }}{{2{{\left ({3b\eta - 2{\theta ^2}} \right) }^2}}}.
	\end{cases}
\end{equation}

\begin{pro}\label{pro1}
	When the investment coefficient $ \eta > \frac{{{\theta ^2}}}{{2b}} $, the game system has an optimal equilibrium solution satisfying the concavity conditions.
\end{pro}

\begin{proof}
	The game model \eqref{eq02} $ \pi_1 $ has the optimal best response strategies if and only if the first order condition holds and the second order derivative is negative, implying strictly concave profit functions. Obviously, $ \frac{\partial^2 \pi_1}{\partial p_1^2} = -2b < 0 $ when $ \frac{\partial \pi_1}{\partial p_1} = 0 $.
	Additionally, the Hessian matrix $ H = \begin{pmatrix}
			{ - 2b}&\theta \\
			\theta &{ - \eta }
	\end{pmatrix} $ needs to be negative definite for the game's optimal equilibrium solution to exist. Obviously, $- 2b < 0 $ alway holds since we have $ 0 < b < 1 $, then let $ \det (H) = 2b\eta - {\theta ^2} < 0 $, it can drive $ \eta < \frac{{{\theta ^2}}}{{2b}} $.
	Thus, we obtain the optimal strategies when $ \eta < \frac{{{\theta ^2}}}{{2b}} $.
\end{proof}

Proposition \ref{pro1} delineates the concavity conditions for the existence of an optimal equilibrium solution in the game model. Based on these sufficient conditions, we can further derive the relationship between pricing of the two data products, as well as compare the relative profits.

\begin{pro}\label{pro2}
	The pricing of the two data products has the following relationship: 
	\begin{enumerate}
		\item $ p_1 > p_2 $ when $ \frac{{{\theta ^2}}}{{2b}} < \eta < \frac{{2{\theta ^2}}}{{3b}} $; 
		\item $ p_1 < p_2 $ when $ \eta > \frac{{2{\theta ^2}}}{{3b}} $.
	\end{enumerate}
\end{pro}
\begin{proof}
	Due to space constraints, the proof is omitted.
\end{proof}

\begin{pro}\label{pro3}
	The profits of the directly sold data product $\pi_1$ and platform-sold data product $\pi_2$ have the following relationships:
	\begin{enumerate}
		\item $ \pi_1 > \pi_2 $ when $\frac{{{\theta ^2}}}{{2b}} < \eta  < \frac{{2{\theta ^2}}}{{3b}}$, $ 0 < r < \frac{{a{\theta ^2}}}{{12{b^2}\eta  - 2b{\theta ^2}}}$, or $\eta  > \frac{{2{\theta ^2}}}{{3b}}$, $r > \frac{{a{\theta ^2}}}{{12{b^2}\eta  - 2b{\theta ^2}}}$; 
		\item $ \pi_1 < \pi_2 $ when $\frac{{{\theta ^2}}}{{2b}} < \eta  < \frac{{2{\theta ^2}}}{{3b}}$, $ r > \frac{{a{\theta ^2}}}{{12{b^2}\eta  - 2b{\theta ^2}}}$, or $\eta  > \frac{{2{\theta ^2}}}{{3b}}$, $0 < r < \frac{{a{\theta ^2}}}{{12{b^2}\eta  - 2b{\theta ^2}}}$.
	\end{enumerate}
\end{pro}
\begin{proof}
	Due to space constraints, the proof is omitted.
\end{proof}

Propositions \ref{pro2} and \ref{pro3} compare the pricing relationship between the two data products under the two distinct data sales models, and contrast profit levels between two enterprises, respectively. As evident in Proposition \ref{pro2}, under the two distinct data sales models, data pricing exhibits associations with several key factors, i.e., the investment coefficient, consumer preference, and cross-price elasticity. When investment coefficient $\eta$ takes relatively lower values ($ \frac{{{\theta ^2}}}{{2b}} < \eta < \frac{{2{\theta ^2}}}{{3b}} $), the pricing for direct enterprise sales $ p_1 $ exceeds that of platform sales $ p_2 $. Conversely, when investment coefficient $\eta$ reaches higher values ($ \eta > \frac{{2{\theta ^2}}}{{3b}} $), the data pricing via the third-party platform $p_2$ surpasses the direct sales price level $p_1$.

\begin{figure}[!htb]
	\centering
		\includegraphics[width = 0.7\linewidth]{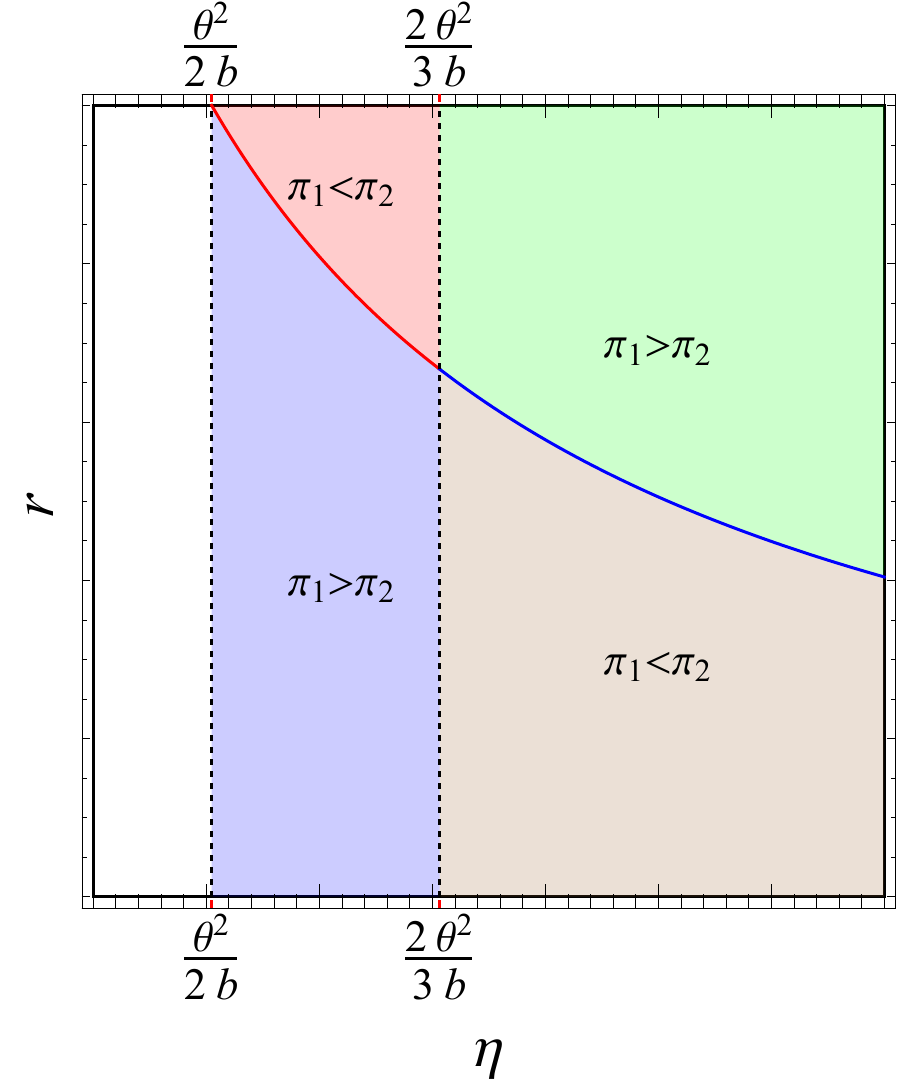}
	\caption{The relationship between $\pi_1$ and $\pi_2$.}
	\label{fig: Fig_pi}
\end{figure}

Comparing the profitability of the two sales models exposes intricate outcomes intertwined with the third-party platform's wholesale pricing $r$ paid to the enterprise.  This pricing parameter directly impacts the profits for both the enterprise and the platform. Firstly, when the wholesale price $r$ takes relatively low values ($ 0 < r < \frac{{a{\theta ^2}}}{{12{b^2}\eta  - 2b{\theta ^2}}}$), if the investment coefficient $\eta$ also assumes lower values ($\frac{{{\theta ^2}}}{{2b}} < \eta  < \frac{{2{\theta ^2}}}{{3b}}$), then $ \pi_1 > \pi_2 $, and vice versa. Conversely, when $r$  higher values, if $\eta$ also higher, then $ \pi_1 < \pi_2 $, and vice versa. The visualized graphical relationship between the profits $\pi_1$ and $\pi_2$ is shown in the Figure \ref{fig: Fig_pi}.

As evident from the preceding Propositions, the investment coefficient $\eta$ reflects the scale of data investment that could determine profits magnitude. 
Counterintuitively, one might presume lower $r$ universally improves platform profitability.   However, given the negligible platform investment willingness despite cheap wholesale costs, the income of the platform may still falter.   This could originate from consumers preferring direct factory purchases for near-homogenous products or perceiving inadequate value in platform offerings.

\section{Dynamic Game Model}\label{sec04}

While the derived single-stage equilibrium solutions offer initial insights, they cannot encapsulate intricacies underpinning repeat interactions. As the game epitomizes dynamic strategic interplays, assuming rational players, both enterprise and platform incorporate information from previous periods in repetitive games. Hence, constructing a dynamic game framework enables studying complex dynamic phenomena and characteristics inherent in long-term strategic behaviors\cite{Ma2022,Liu2022}.

As competitors' strategies remain partially obscured over successive decisions, game players follow bounded rationality in periodically adjusting approaches.   Specifically, at each period $t$ $(t = 1, 2, \cdots , n)$, decisions for the upcoming period $t+1$ adapt based on historically revealed information.   Therein, the dynamic centralized decision-making model evolves as follows:
\begin{equation}\label{eq05}
	\begin{cases}
		p_1 (t+1) = p_1 (t) + \alpha_1 p_1 (t) \frac{{\partial {\pi _1 (t) }}}{{\partial {p_1 (t) }}}\\
		p_2 (t+1) = p_2 (t) + \alpha_2 p_2 (t) \frac{{\partial {\pi _2 (t) }}}{{\partial {p_2 (t) }}}\\
		s (t+1) = s (t) + \alpha_3 s (t) \frac{{\partial {\pi _2 (t) }}}{{\partial s (t) }}
	\end{cases}
\end{equation}
where $\alpha_i~ (i=1,2,3)$ denotes the adjustment rate of strategy changes for each player, encapsulating decision-makers' responsiveness to market evolutions. Moreover, imposing non-negativity constraints on decision variables ensures practical relevance aligned with realistic production scenarios. 

Solving the dynamic game system \eqref{eq05} derivation yields eight equilibria which are listed in Table \ref{tab02}. Imposing non-negativity constraints, as strategies bearing no enterprise value prove unrealistic\footnote{While negative strategy equilibria are mathematically valid solutions, they denote instability in the modeled gaming system, devoid of real-world practicality.  Hence, selectively constraining the feasible space to positive pricing strategies not only filters but also distinctly identifies the equilibrium outcome aligned with reality-consistent implementation constraints.}, so $E^* = E_8$ is the only dynamically stable equilibrium solution.

\begin{table}[!htb]
	\centering
	\caption{Eight equilibrium solutions of dynamic system \eqref{eq05}.}\label{tab02}
	\begin{tabular}{cccc}
		\toprule
		$E_i$ & $p_1$ & $p_2$ & $s$ \\
		\midrule
		$E_1$ &	0&0&0\\
		$E_2$ &	0&0& $- \frac{{r\theta }}{\eta }$\\
		$E_3$ &	0&$\frac{{a + br}}{{2b}}$&0\\
		$E_4$ &	0&$r + \frac{{ (a - br)\eta }}{{ 2b\eta - {\theta ^2}}}$&$\frac{{\left ({a - br} \right) \theta }}{{2b\eta - {\theta ^2}}}$\\
		$E_5$ &	${\frac{a}{b} + r}$&$\frac{a}{b} + r$&0\\
		$E_6$ &	$\frac{{a + br}}{{2b}}$&0&0\\
		$E_7$ &	$\frac{{a + br}}{{2b}}$&0& $- \frac{{r\theta }}{\eta }$\\
		$E_8$ &	$r + \frac{1}{2}a\left ({\frac{1}{b} + \frac{{3\eta }}{{3b\eta - 2{\theta ^2}}}} \right)$ &$r + \frac{{3a\eta }}{{3b\eta - 2{\theta ^2}}}$&$\frac{{3a\theta }}{{3b\eta - 2{\theta ^2}}}$\\
		\bottomrule
	\end{tabular}	 
\end{table}

\begin{pro}\label{pro4}
The Nash equilibrium decision $E^*$ is locally asymptotically stable when constraint \eqref{eq06} is met.
\begin{equation}\label{eq06}
	\begin{cases}
		1 + a_2 + a_1 + a_0 > 0 \\
		1 -a_2 + a_1 - a_0 > 0 \\
		1 > | a_0 | \\
		| 1 - a_0^2 | > | a_1 - a_0 a_2 |
	\end{cases}
\end{equation}

\end{pro}

\begin{proof}
	The Jacobian matrix of the dynamic system \eqref{eq05} evaluated at the Nash equilibrium point $E^*$ is denoted by $J(p_1^*, p_2^*, s^*)$.
	\begin{equation}\label{eq07}
		J = \begin{pmatrix}
			j_{11}&{b{p_1}{\alpha _1}}&0\\
			{b{p_2}{\alpha _2}}&j_{22}&{{p_2}{\alpha _2}\theta }\\
			0&{s{\alpha _3}\theta }&j_{33}
		\end{pmatrix}
	\end{equation}
	where $j_{11} = 1 + {\alpha _1}( a + b( {p_2} - 4{p_1} + r ) ) $, $j_{22} = 1 + {\alpha _2}( a + b( {p_1} - 4{p_2} + r) + s\theta  )$, $j_{33} = 1 - {\alpha _3}( 2s\eta  - \left( {{p_2} - r} \right)\theta  ) $.
	
	Thus, the characteristic polynomial of Jacobian matrix $J(p_1^* , p_2^*, s^*)$ is denoted by $P (\lambda) = \lambda_3 + a_2 \lambda_2 + a_1 \lambda + a_0$ , with $a_2, a_1, a_0 $ as coefficient terms. The Jury stability criteria applied to this cubic polynomial determines stability of the dynamic system \eqref{eq05} at Nash equilibrium point $E^*$.
\end{proof}

Proposition \ref{pro4} delineates stability conditions for the long-term strategies of the energy enterprise and platform under constraint \eqref{eq06}. Fulfilling this constraint enables decision-makers to calibrate parameters within ranges upholding equilibrium stability. Violating the stability region nevertheless precipitates intricate dynamical phenomena like bifurcations, chaos, and multi-stability. Therefore, we next analyze local and global bifurcation properties to elucidate the origins underpinning these multifaceted complex characteristics.

\subsection{Local Bifurcation}

In long-run dynamic gaming, decision-makers predominantly calibrate optimal strategies by adjusting parameters.   Therefore, analyzing the effects of adjustment parameters on the localized stability of the dynamic system is important.

\begin{figure}[!htb]
	\centering
		\includegraphics[width = 0.7\linewidth]{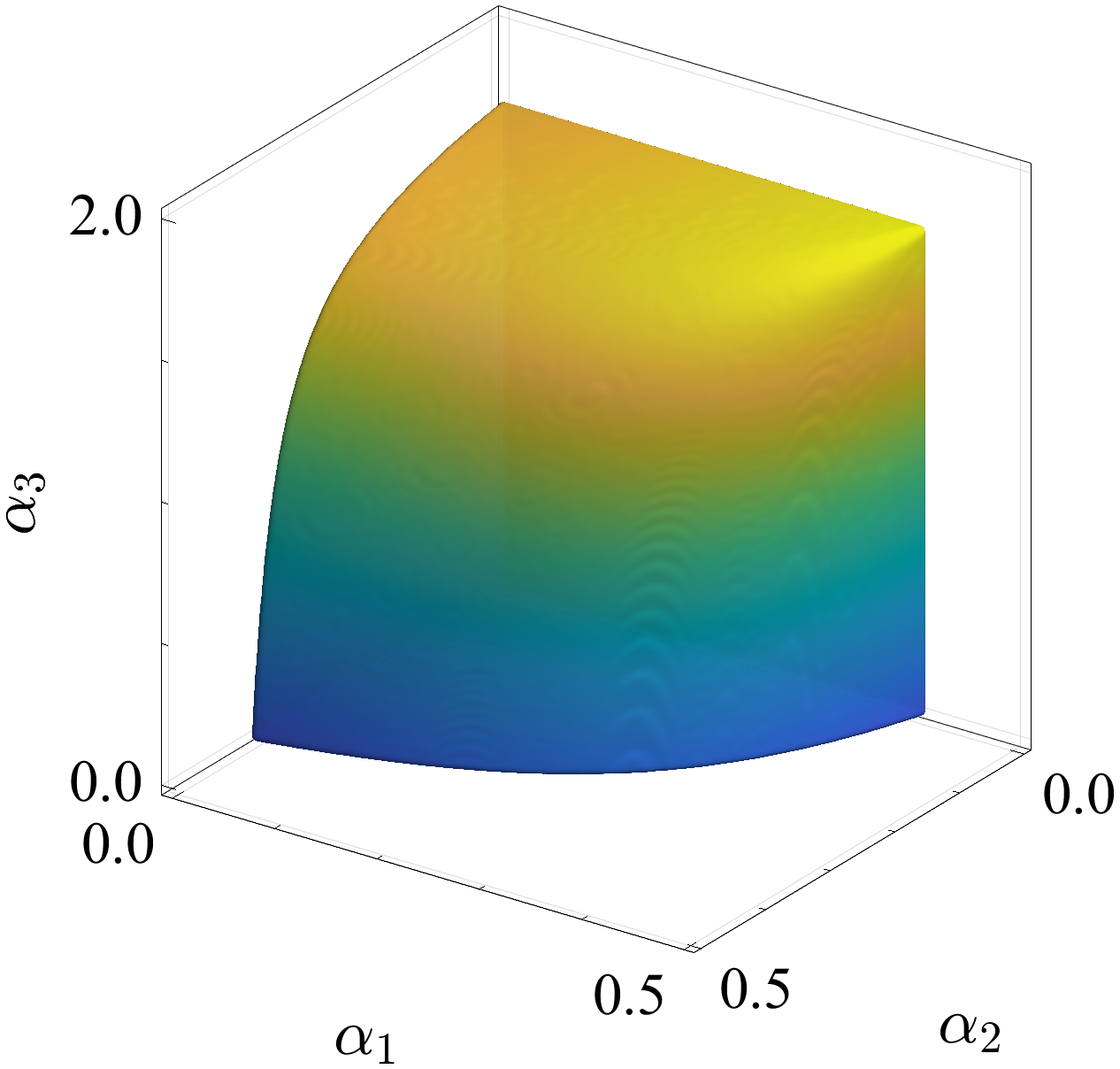}
	\caption{Stable region in $(\alpha_1, \alpha_2, \alpha_3)$ space.}
	\label{fig: Fig_stable3}
\end{figure}

To better analyze the local bifurcation, we can explore the effect of adjustment parameters on dynamic system \eqref{eq05} by numerical simulation. Fix the parameter values: $a = 2, b = 0.4, \theta = 0.2, \eta = 0.5, r = 0.6 $ and consider the constraint \eqref{eq06}, we can illustrate the stable region in Figure \ref{fig: Fig_stable3}, which shows the significant impact of adjustment parameters on the dynamic system \eqref{eq05}.

\begin{figure}[!htb]
	\centering
	\subfloat[Bifurcation diagram with varing $\alpha_1$.\label{fig: Fig_bif_alpha}]{\includegraphics[width = 0.6\linewidth]{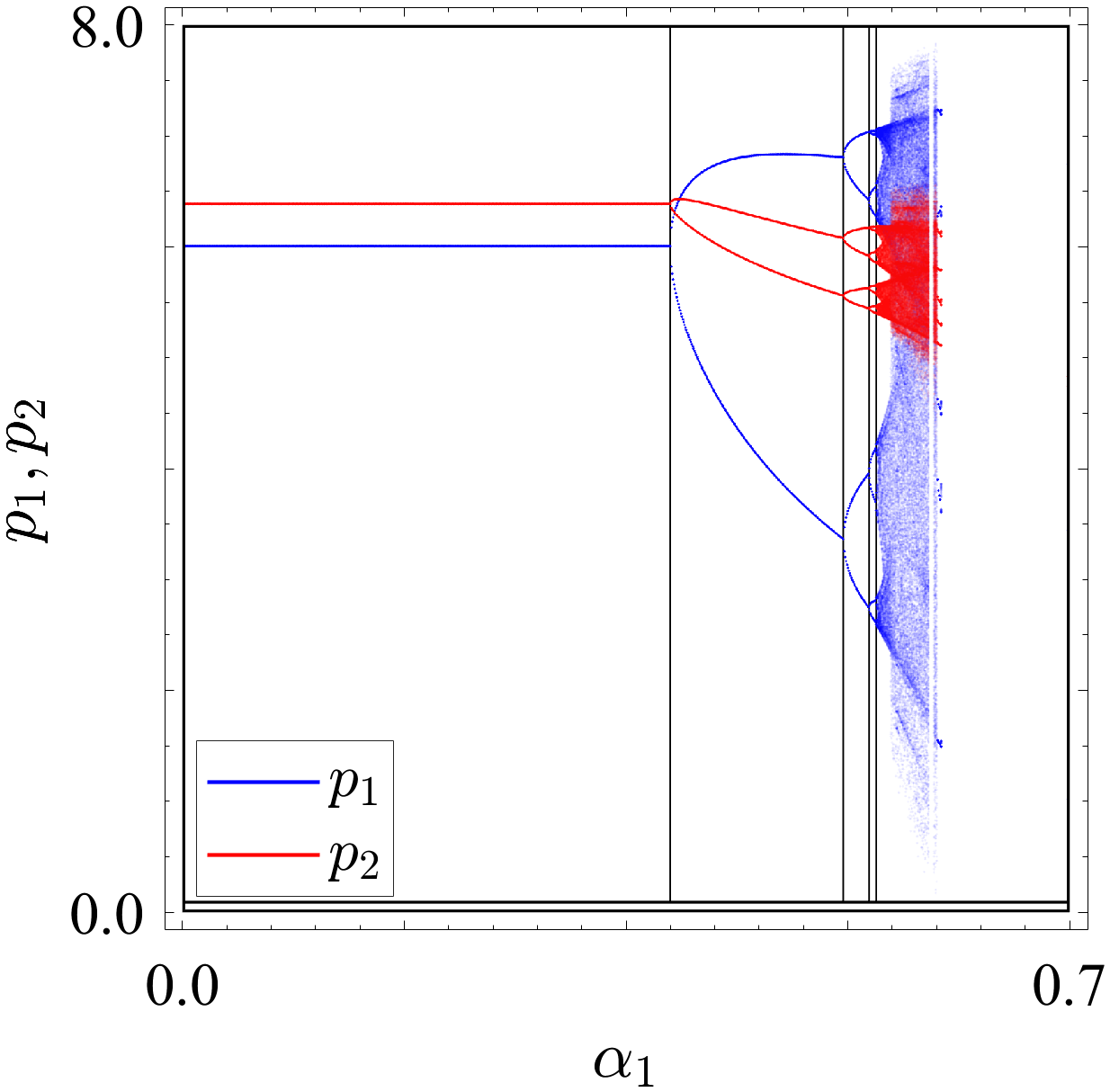}}\\
	\subfloat[Largest Lyapunov exponent diagram with varing $\alpha_1$.\label{fig: Fig_LLE_alpha}]{\includegraphics[width = 0.6\linewidth]{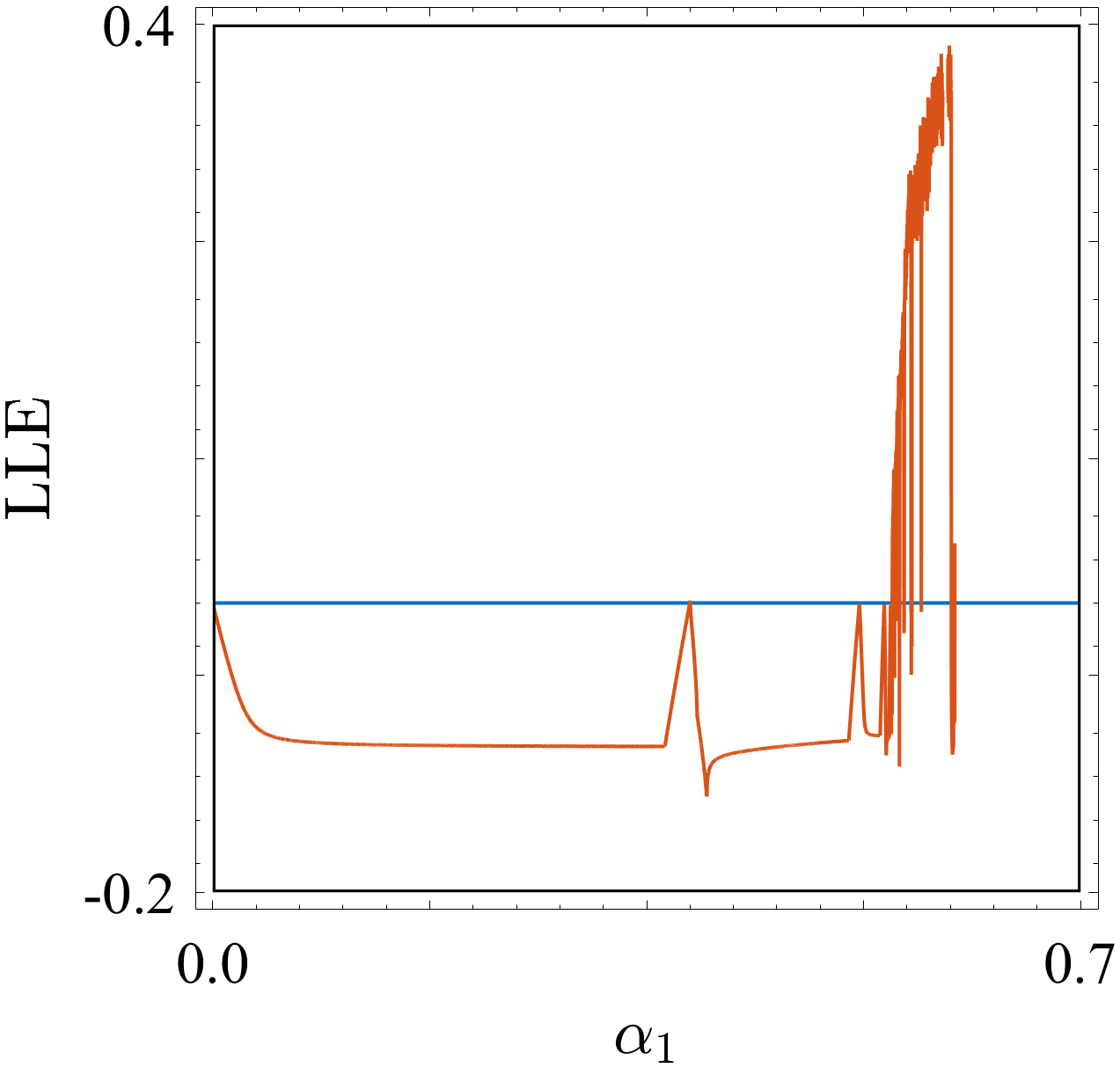}}	
	\caption{Bifurcation and largest Lyapunov exponent diagrams of system \eqref{eq05}.}
	\label{fig: Fig_bif_LLE}
\end{figure}

Figure \ref{fig: Fig_stable3} delineates the stability region whereby tuning parameters within this domain. Complex dynamics behaviors may emerge when parameters are outside the domain. Figure \ref{fig: Fig_bif_alpha} exhibits how varying adjustment rate $\alpha_1$ gradually destabilizes the system's equilibrium into period-doubling bifurcations, eventually chaos. Such unpredictable, disordered equilibria adversely impact market efficiency and competition. Correspondingly, Figure \ref{fig: Fig_LLE_alpha} plots the Largest Lyapunov Exponent across different $\alpha_1$, the positive values indicate chaos while non-positive values signify stability. This metric visually communicates complex shifts as parameters change, guiding decision-makers adjustments. Similarly, varying the other adjustment parameters ($\alpha_2, \alpha_3 $) elicits analogous complex dynamical phenomena that depicted above.

\begin{figure*}[!htb]
	\centering
	\subfloat[\label{fig: figbasina}]{\includegraphics[height = 0.15\linewidth]{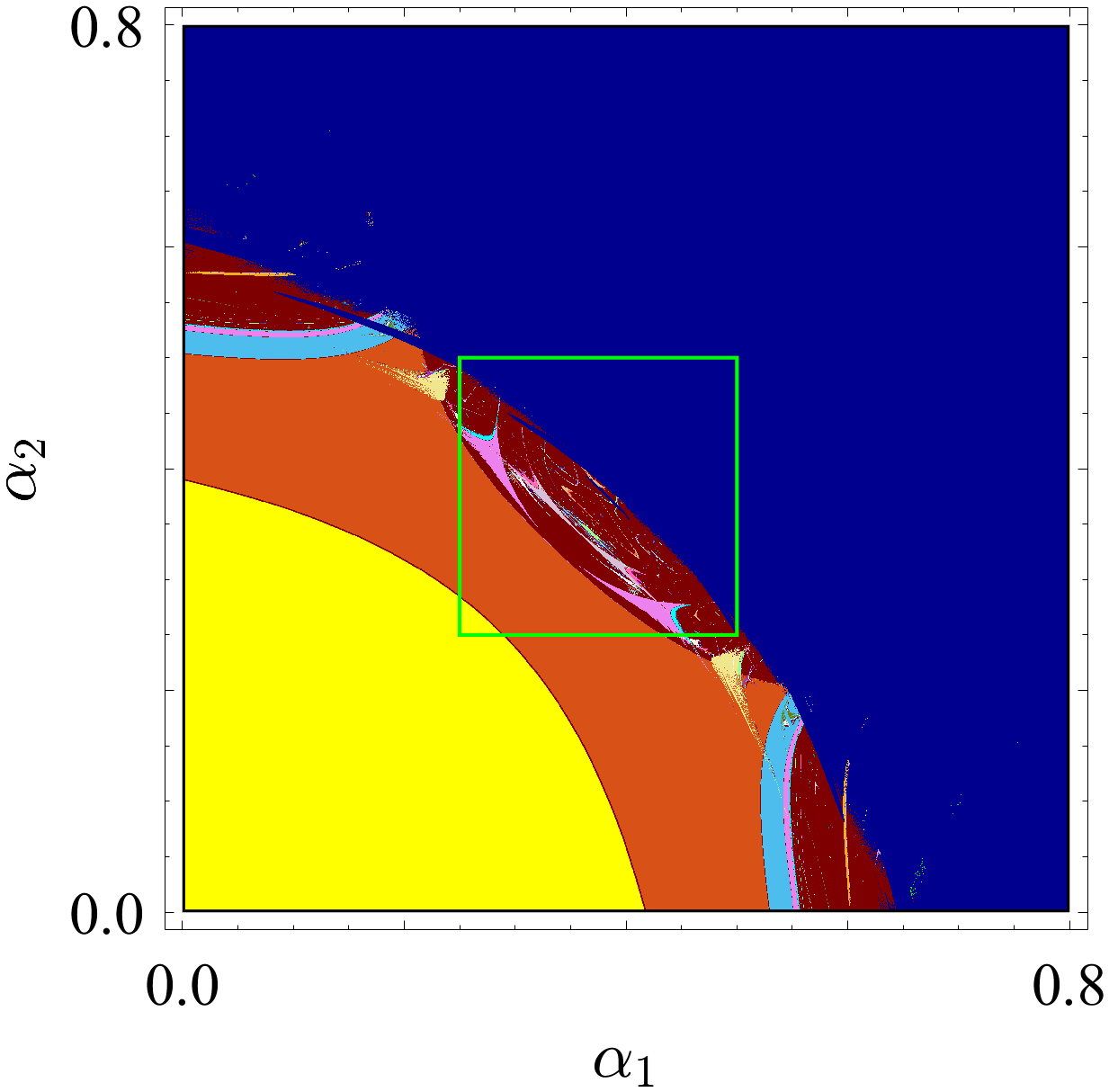}} \hfill
	\subfloat[\label{fig: figbasinb}]{\includegraphics[height = 0.15\linewidth]{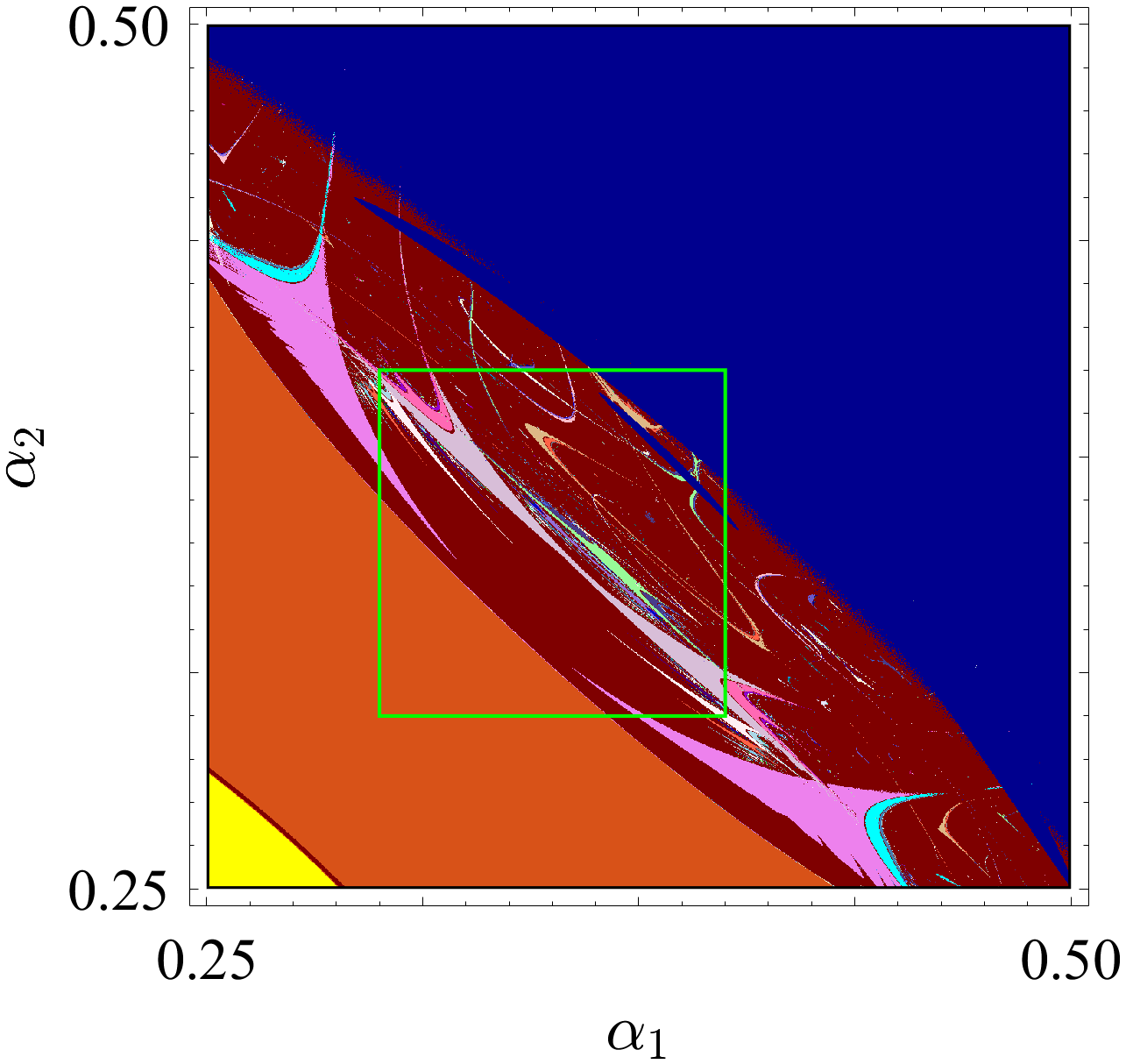}} \hfill
	\subfloat[\label{fig: figbasinc}]{\includegraphics[height = 0.15\linewidth]{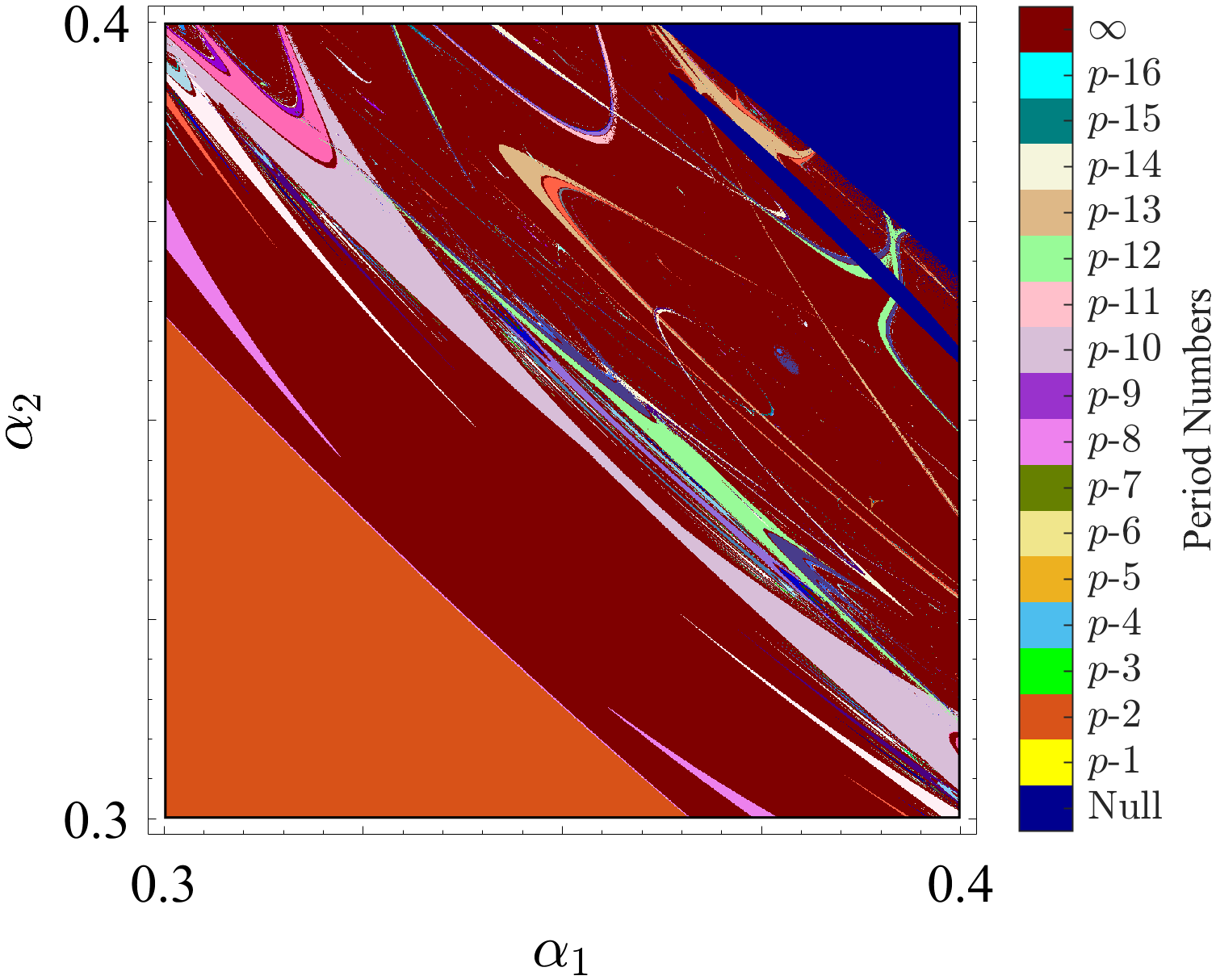}}	\hfill
	\subfloat[\label{fig: figbasind}]{\includegraphics[height = 0.15\linewidth]{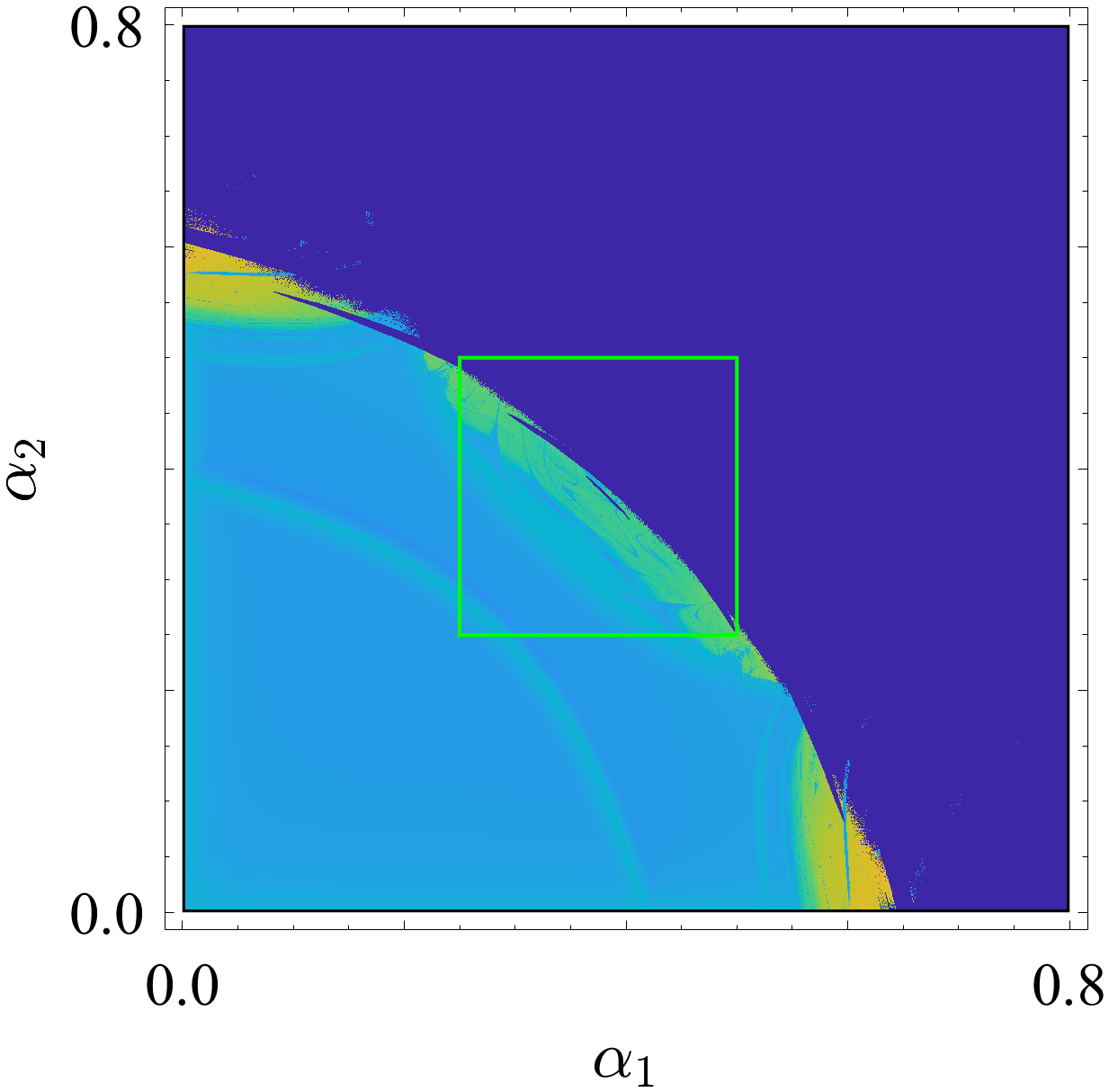}}\hfill
	\subfloat[\label{fig: figbasine}]{\includegraphics[height = 0.15\linewidth]{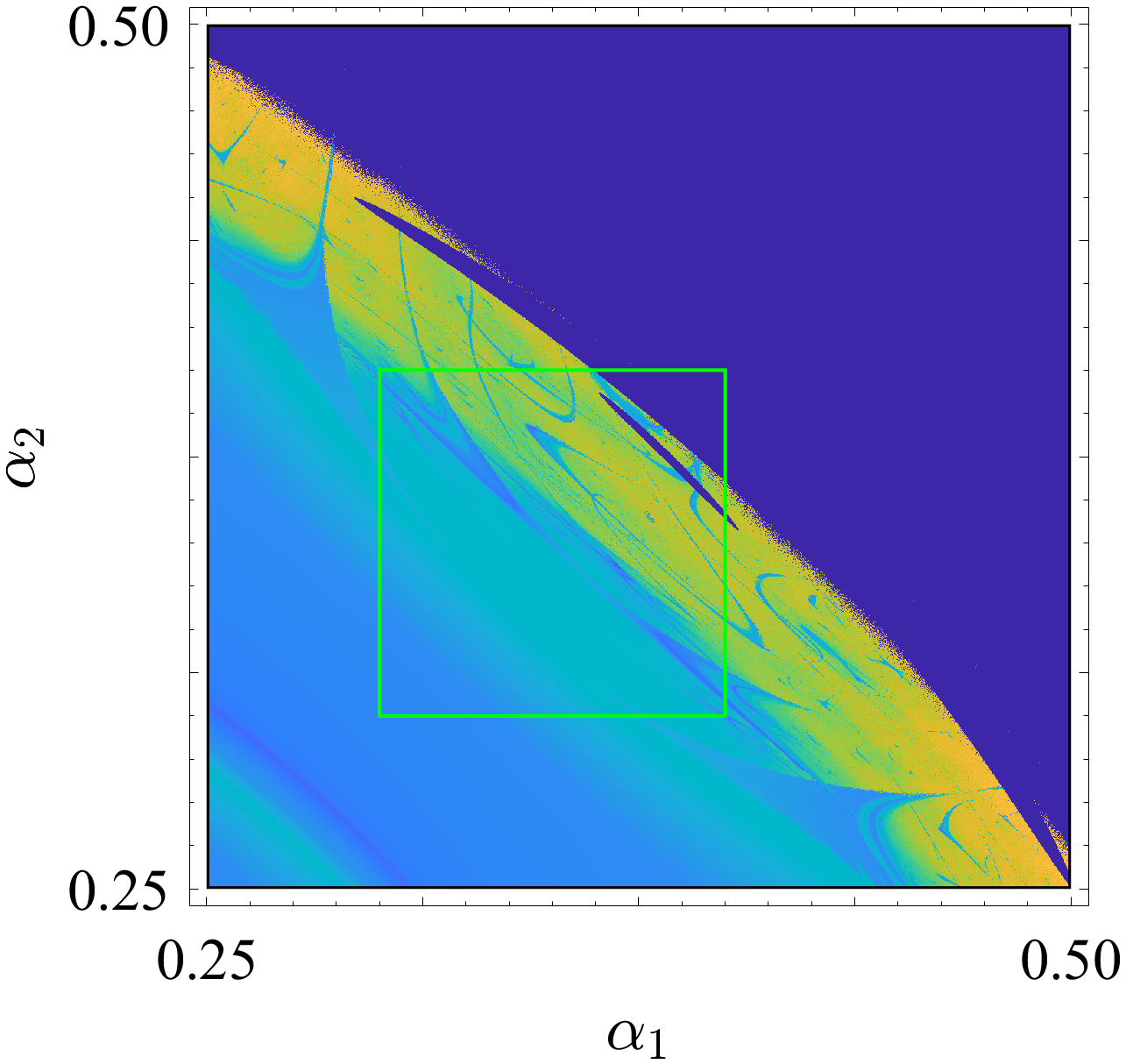}}\hfill
	\subfloat[\label{fig: figbasinf}]{\includegraphics[height = 0.15\linewidth]{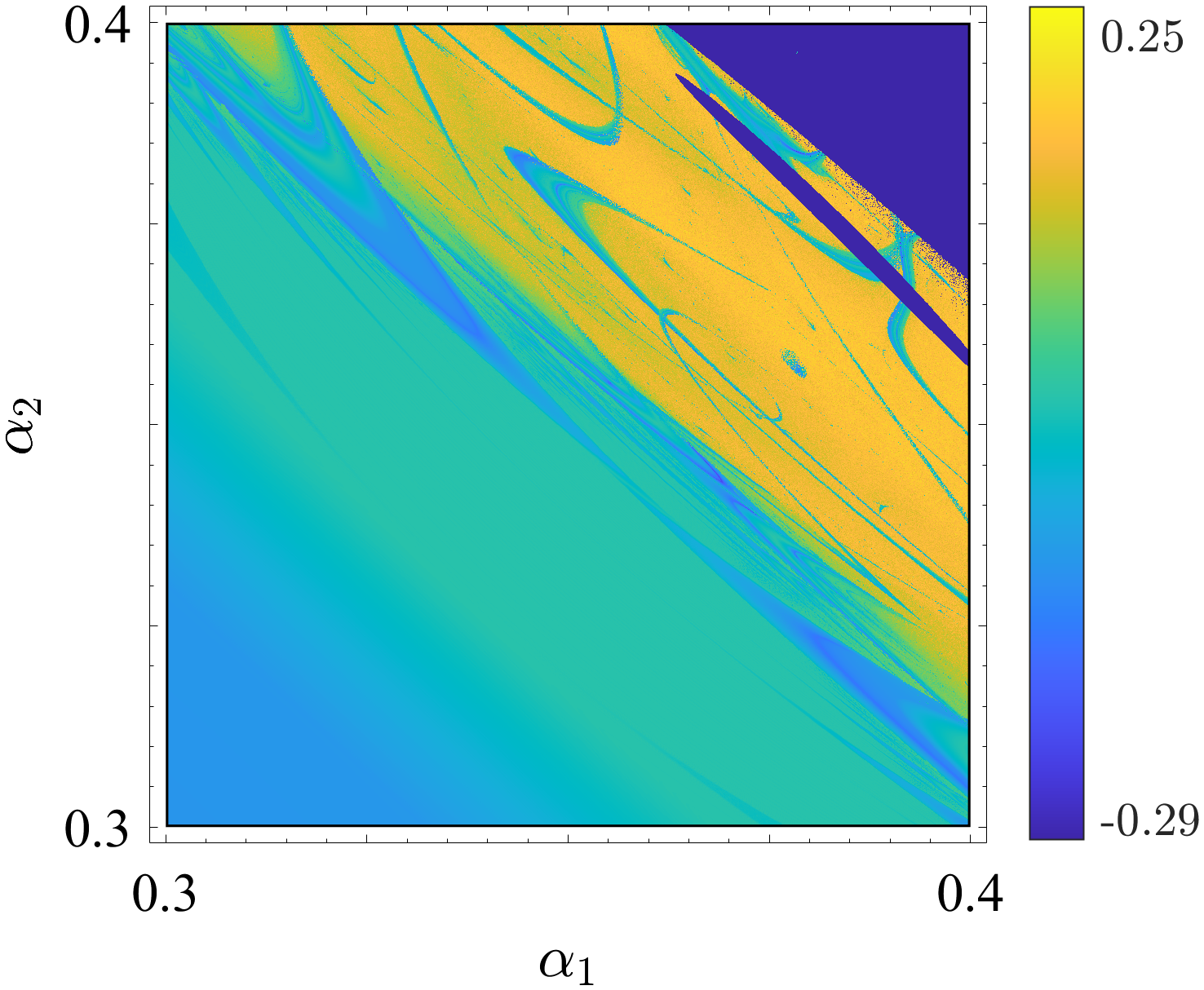}}
	\caption{Bifurcation diagram and largest Lyapunov exponent diagram on the parameter plane of $(\alpha_1, \alpha_2)$.}
	\label{fig: figbasin}
\end{figure*}

While varying individual parameters elicit complex bifurcations, simultaneous multi-parameter adjustments further compound complexity.   Demonstrating this phenomenon, Figure \ref{fig: figbasin} portrays two-dimensional bifurcation and the Largest Lyapunov Exponent across $\alpha_1$ and $\alpha_2$.  It reveals interdependent impacts from concurrent parameter changes.

Figure \ref{fig: figbasin} shows two-dimensional bifurcation and largest Lyapunov exponent diagram on the $(\alpha_1, \alpha_2)$ plane, with color-coded regions denoting distinct periodic orbits and levels of stability. Specifically, Figure \ref{fig: figbasina} - \ref{fig: figbasinc} are two-dimensional bifurcation diagrams, the yellow region signifies stable fixed point solutions; orange region designates stable period-two cycles; light blue region indicates stable period-four cycles; while dark red and blue respectively signify chaotic and divergent zones. Traversing this dynamical landscape reveals that subtle combined parameter variations can profoundly destabilize local equilibria – flipping stability into unpredictable chaos. Hence, cautiously navigating the periphery and understanding interdependencies proves critical when adjusting multiple parameters simultaneously.
Figure \ref{fig: figbasind} - \ref{fig: figbasinf} are two-dimensional largest Lyapunov exponent diagram, with deep blue signifying divergent domain. Additionally, the color gradation from blue towards yellow indicates the transition of largest Lyapunov exponent from negative into positive territory - portending the onset of chaos. 
As we analyze the stability landscapes, we clearly see that simultaneously tuning multiple key parameters can profoundly transform the system dynamics in complex ways – either driving it towards equilibrium or chaos unpredictability.    Whether enhancing stability or elevating disorder, the interplay across parameter changes dramatically alters outcomes.    Therefore, decision-makers must very carefully, judiciously, and skillfully manage the navigation of parameter combinations based on their interdependencies.    This proves critical in charting an optimal path forward amidst complex terrain.    Rather than adjusting blindly, understanding interactive effects allows decisions movement toward secure results.

In summary, even small concurrent parameter changes deeply intertwine, necessitating a roadmap to reach a stable state for how adjustments impact one another in reshaping the dynamics that unfold.

\subsection{Global Stable}

Preceding part of this paper analysis explored local bifurcation properties by examining critical parameter impacts on dynamic game model stability. However, alternate initial strategies may also converge towards radically distinct outcomes. Because initial states profoundly shape dynamic systems, investigating global bifurcations through basin and attractor concepts are critical. Specifically, attractor and basins of attraction elucidate macroscopic long-term behaviors - unveiling multistability and exposing sensitivity where minor perturbations near critical thresholds trigger transitions between complex regimes. Therefore, this paper examines global bifurcation patterns combined with attractor and basin of attraction.
\begin{figure*}[!htb]
	\centering
	\subfloat[\label{fig: Fig_basina}]{\includegraphics[height = 0.16\linewidth]{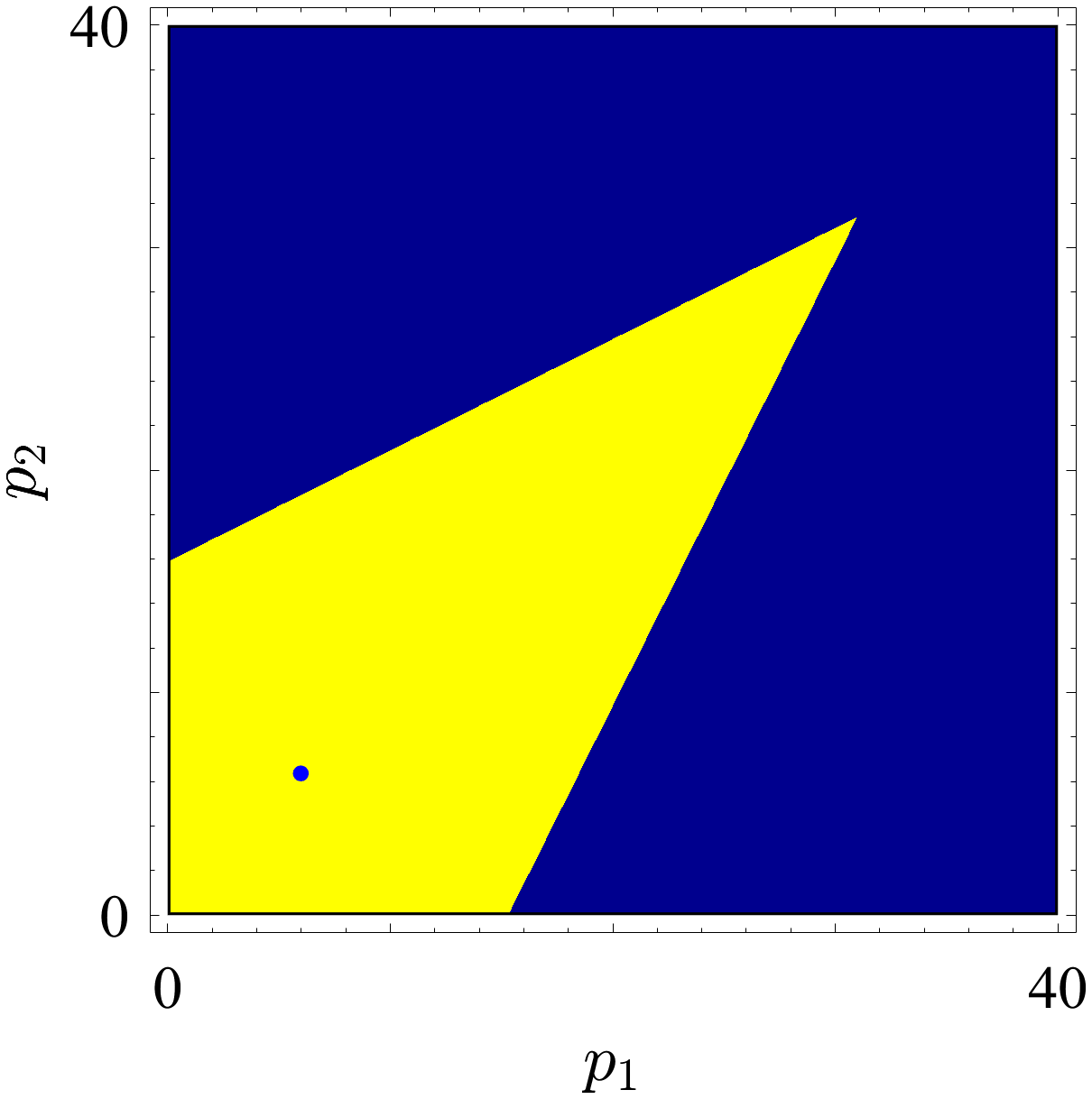}}\hfill
	\subfloat[\label{fig: Fig_basinb}]{\includegraphics[height = 0.16\linewidth]{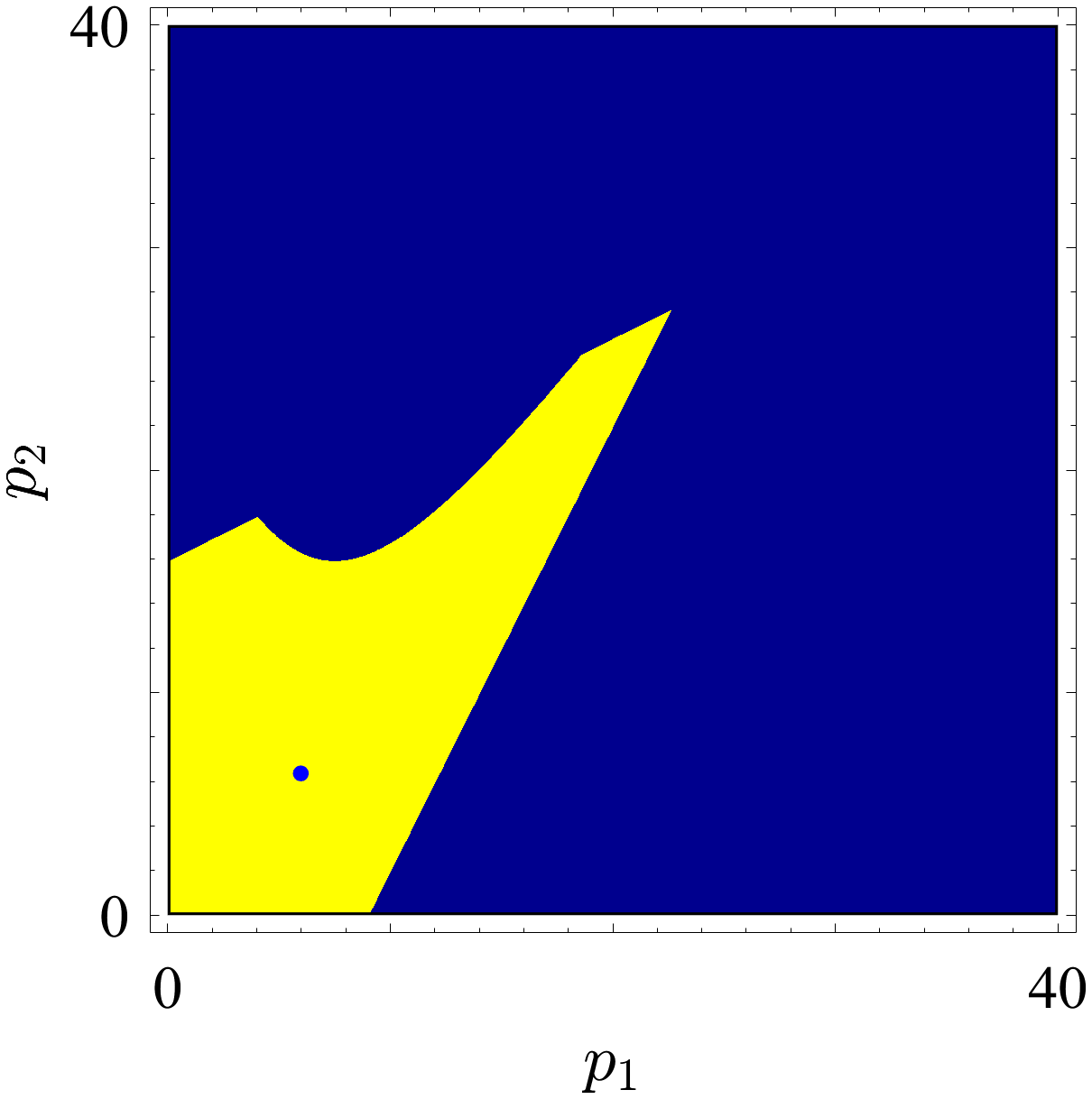}}\hfill
	\subfloat[\label{fig: Fig_basinc}]{\includegraphics[height = 0.16\linewidth]{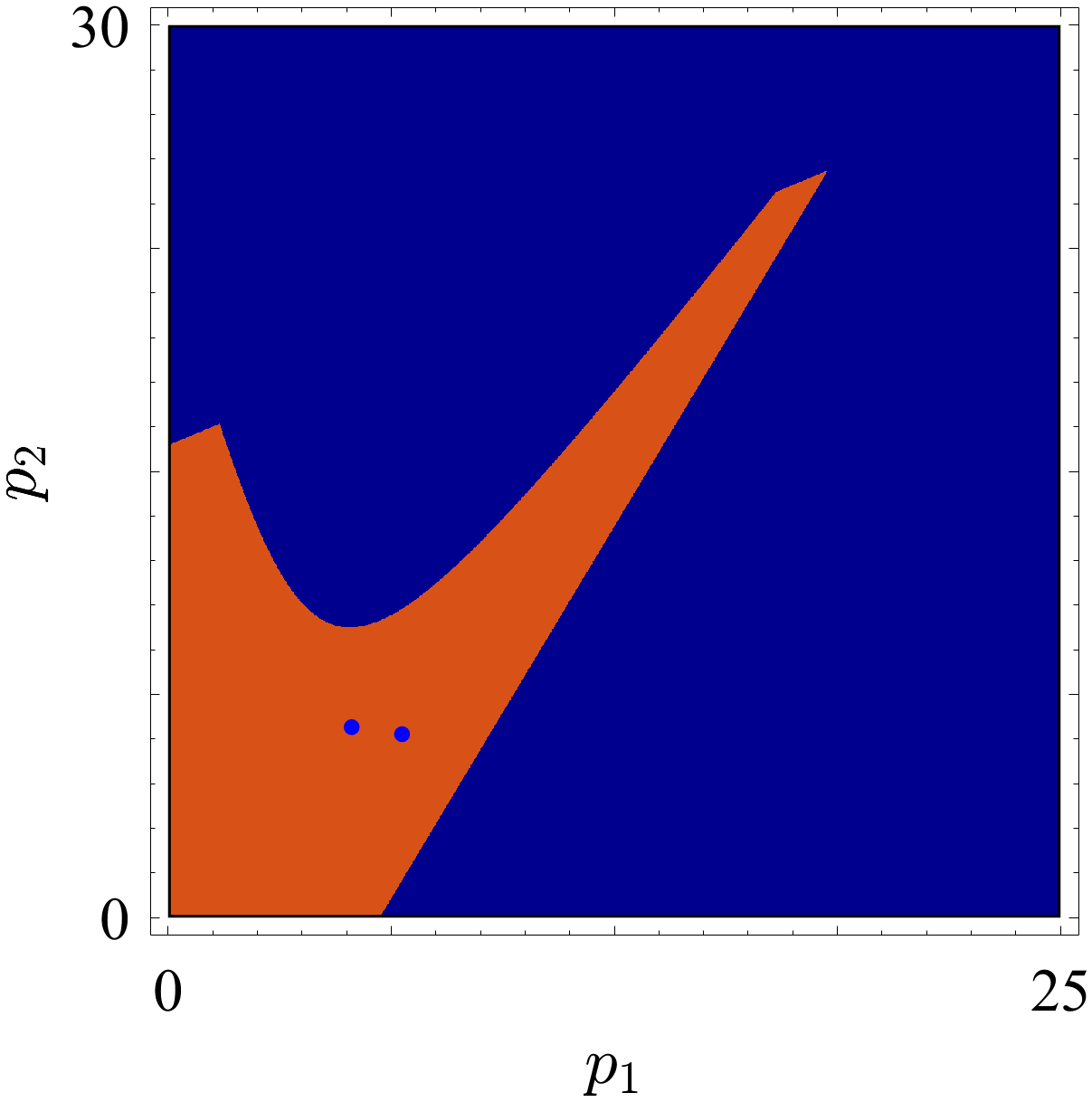}}\hfill
	\subfloat[\label{fig: Fig_basind}]{\includegraphics[height = 0.16\linewidth]{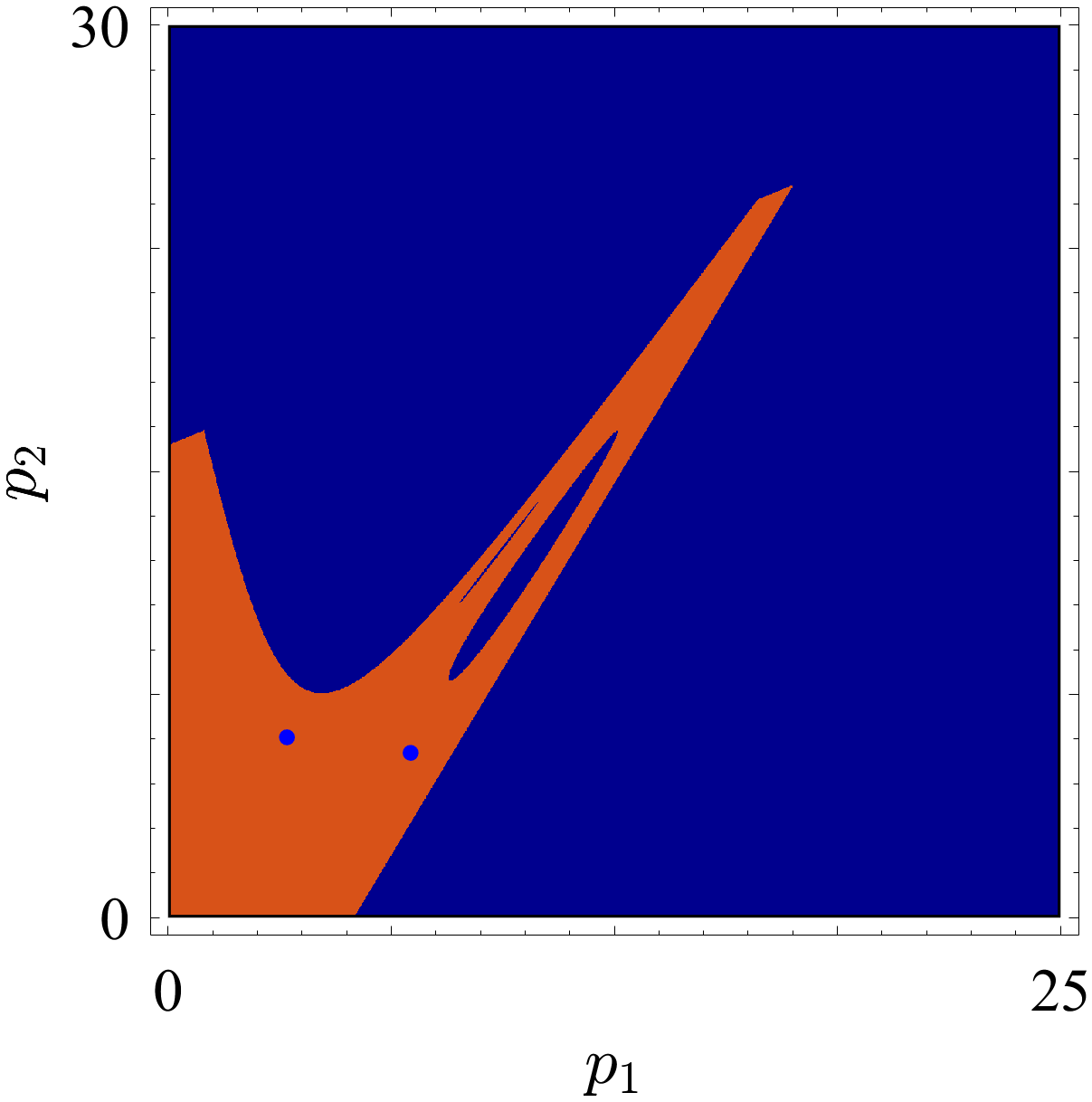}}\hfill
	\subfloat[\label{fig: Fig_basine}]{\includegraphics[height = 0.16\linewidth]{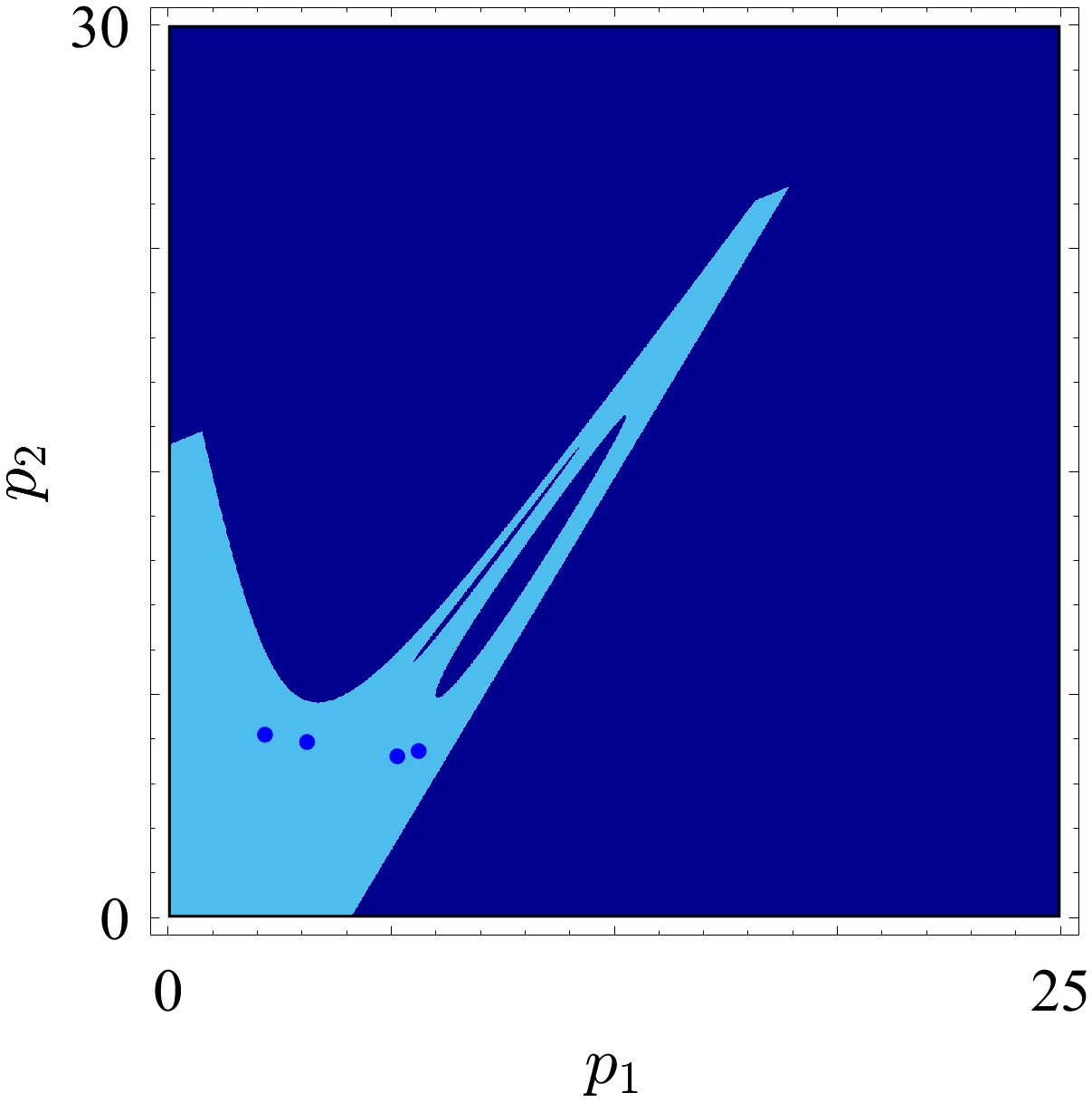}}\hfill
	\subfloat[\label{fig: Fig_basinf}]{\includegraphics[height = 0.16\linewidth]{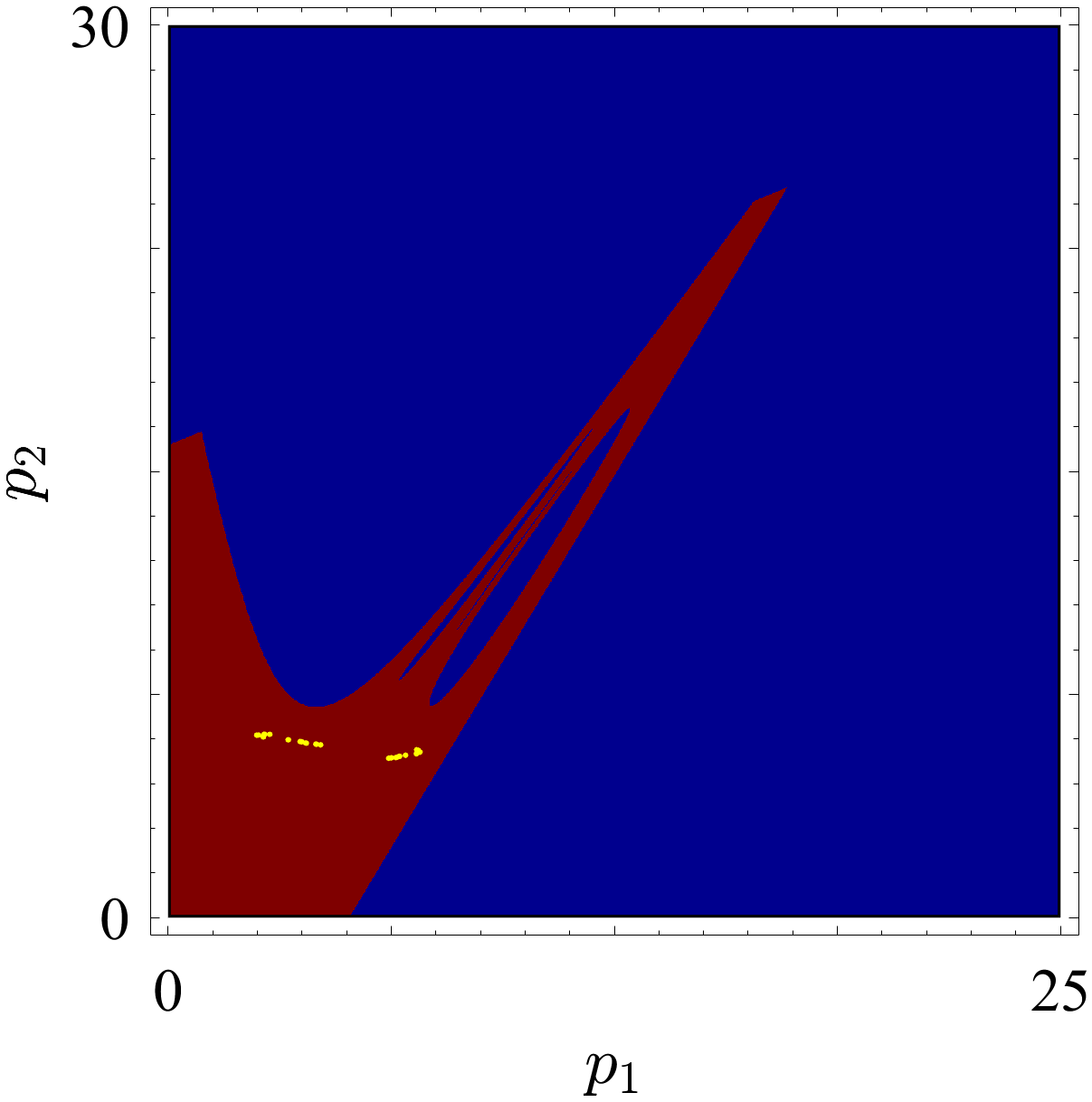}}\\
	\caption{Basins of attraction and projection of attractors in the $(p_1, p_2)$ decision space.}
	\label{fig: Fig_basin1}
\end{figure*}
\begin{figure*}[!htb]
	\centering
	\subfloat[\label{fig: Fig_basing}]{\includegraphics[height = 0.16\linewidth]{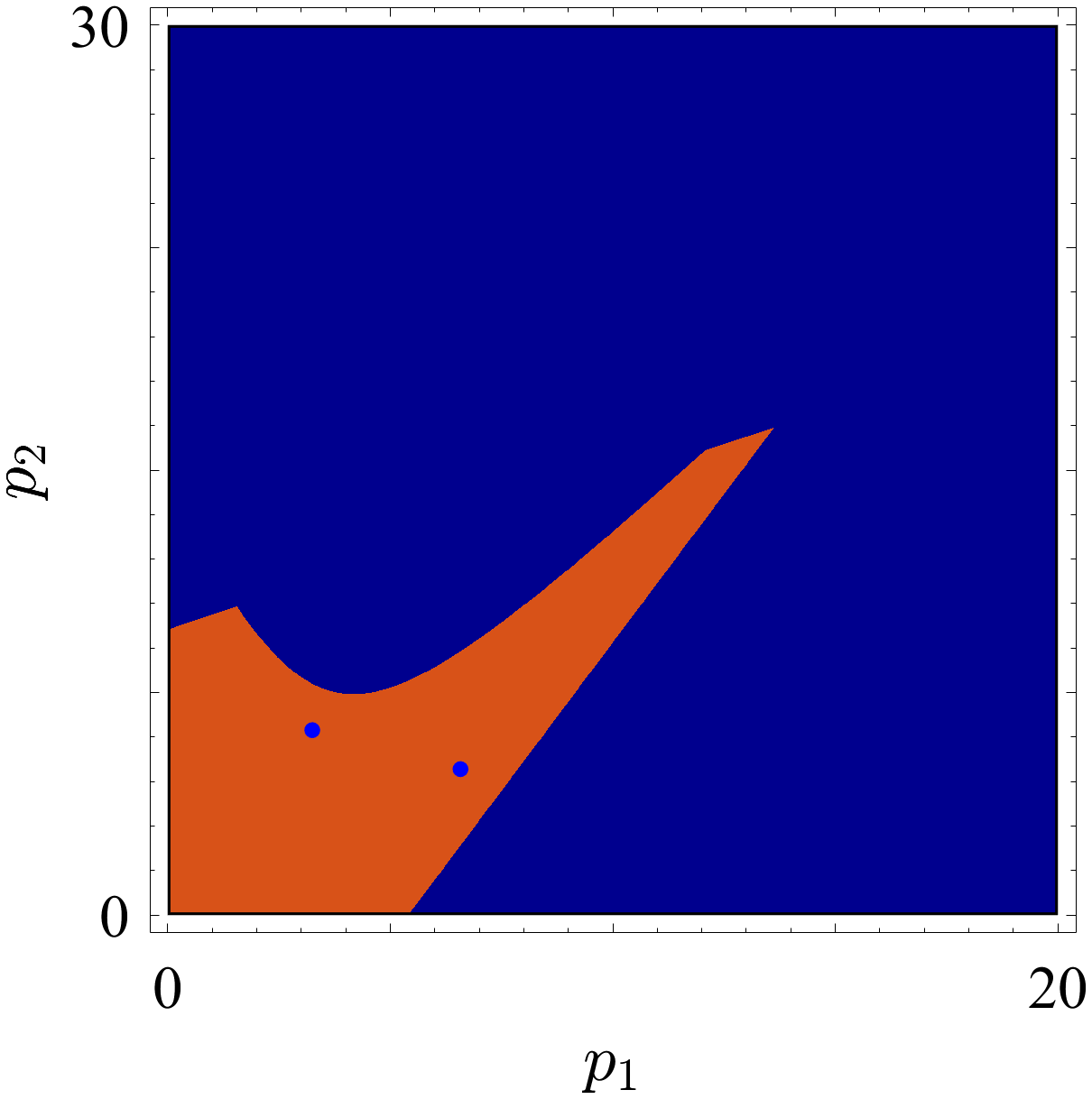}}\hfill
	\subfloat[\label{fig: Fig_basinh}]{\includegraphics[height = 0.16\linewidth]{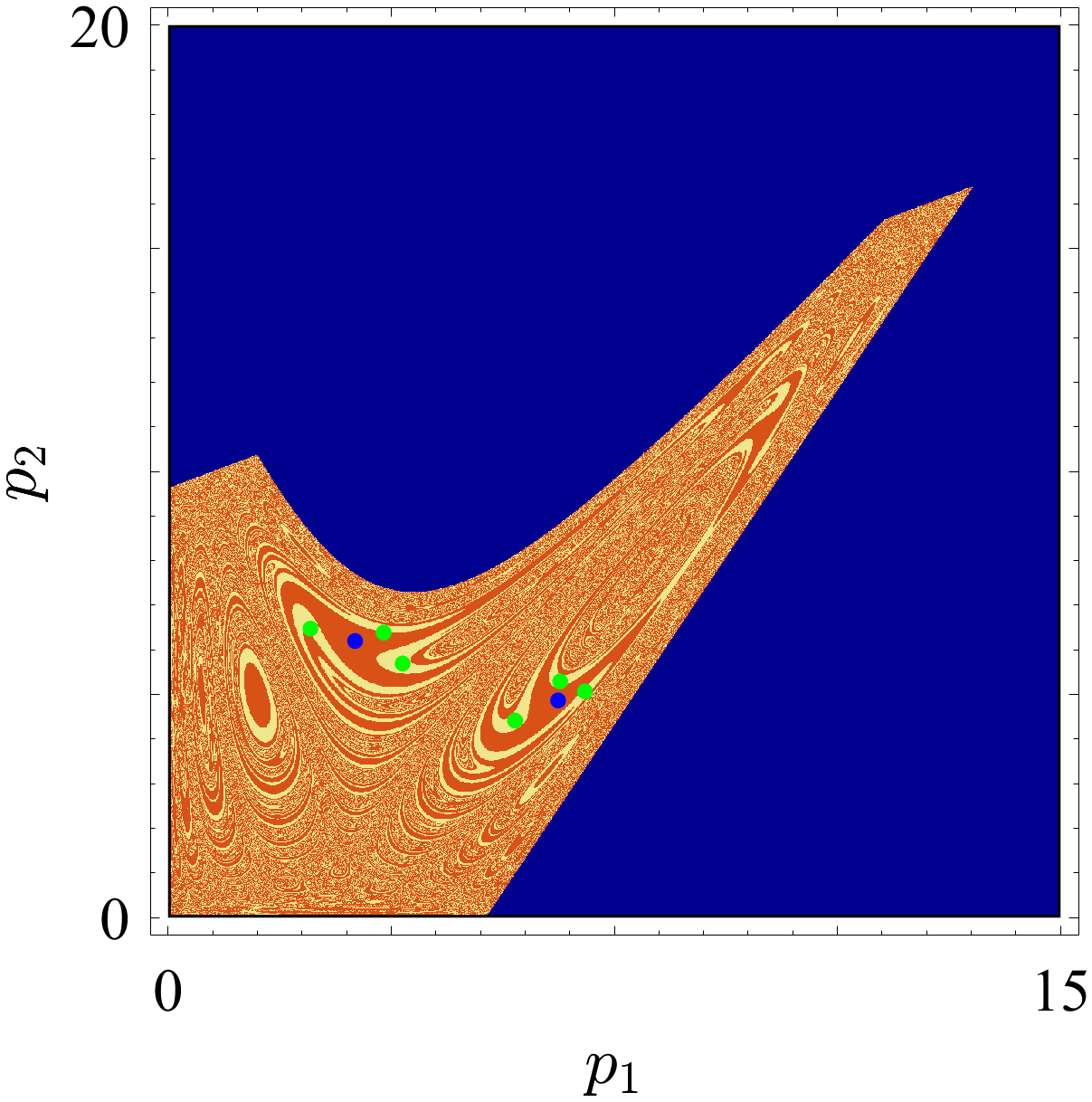}}\hfill
	\subfloat[\label{fig: Fig_basinj}]{\includegraphics[height = 0.16\linewidth]{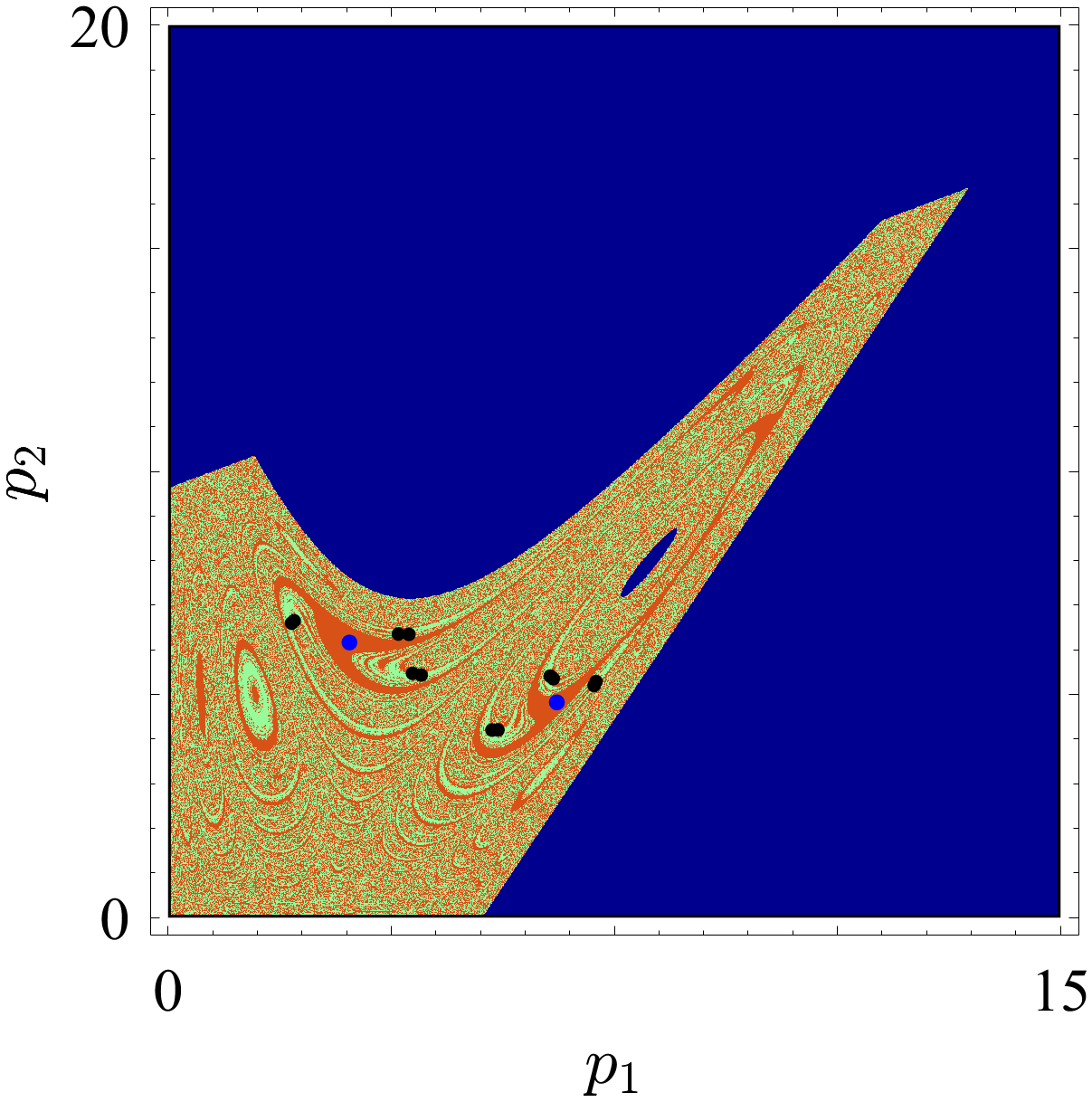}}\hfill
	\subfloat[\label{fig: Fig_basink}]{\includegraphics[height = 0.16\linewidth]{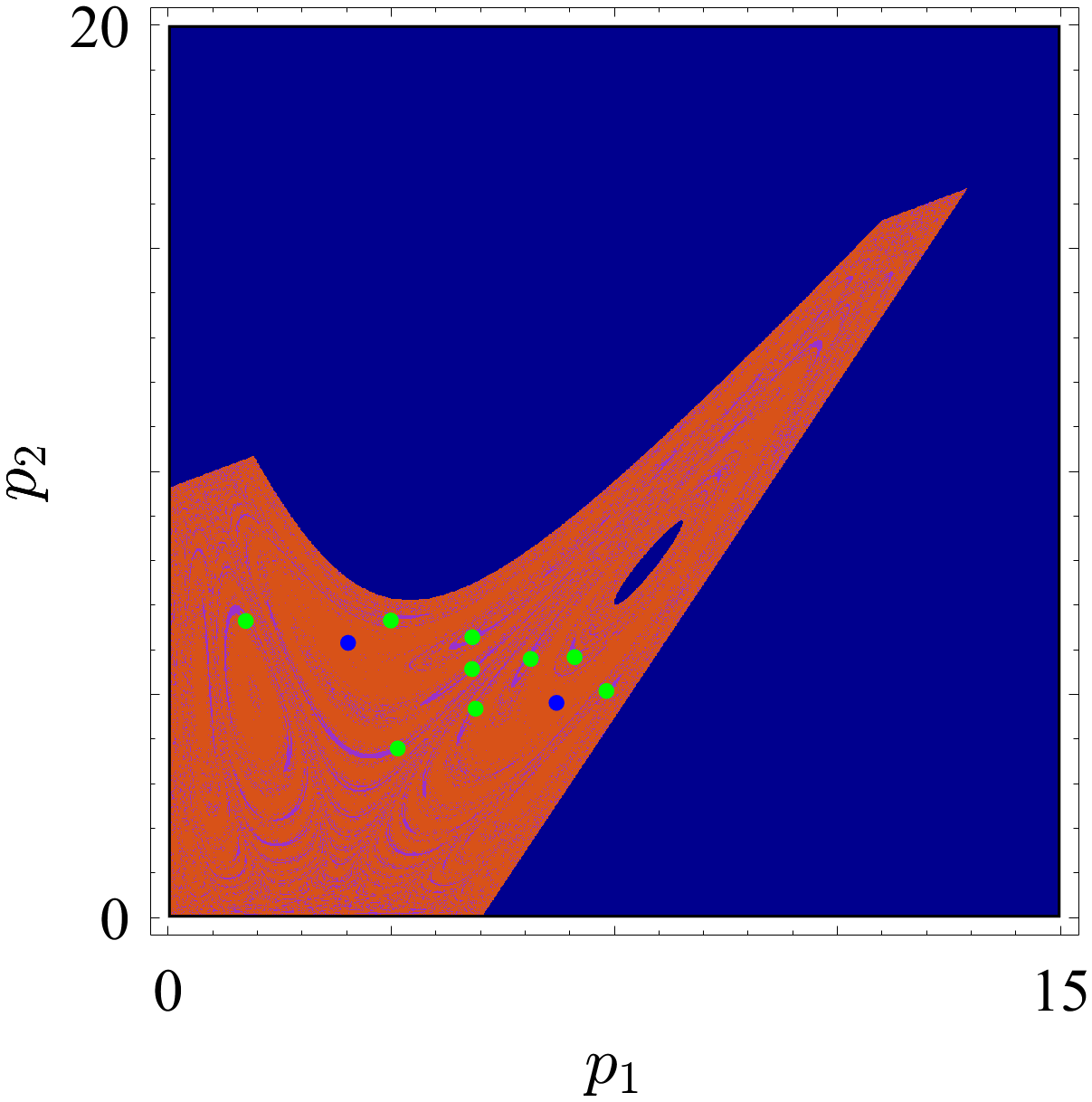}}\hfill
	\subfloat[\label{fig: Fig_basinl}]{\includegraphics[height = 0.16\linewidth]{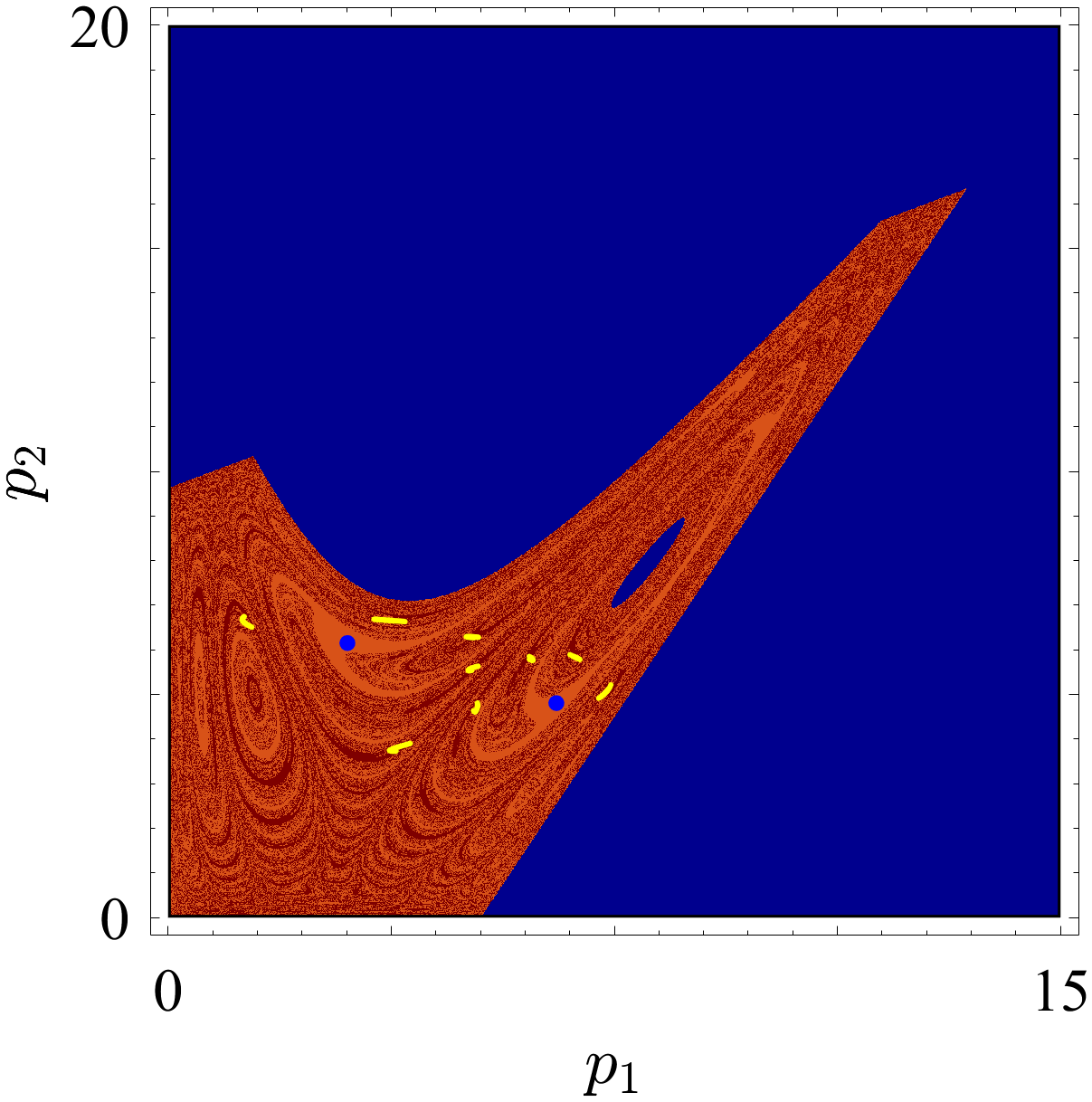}}\hfill
	\subfloat[\label{fig: Fig_basinm}]{\includegraphics[height = 0.16\linewidth]{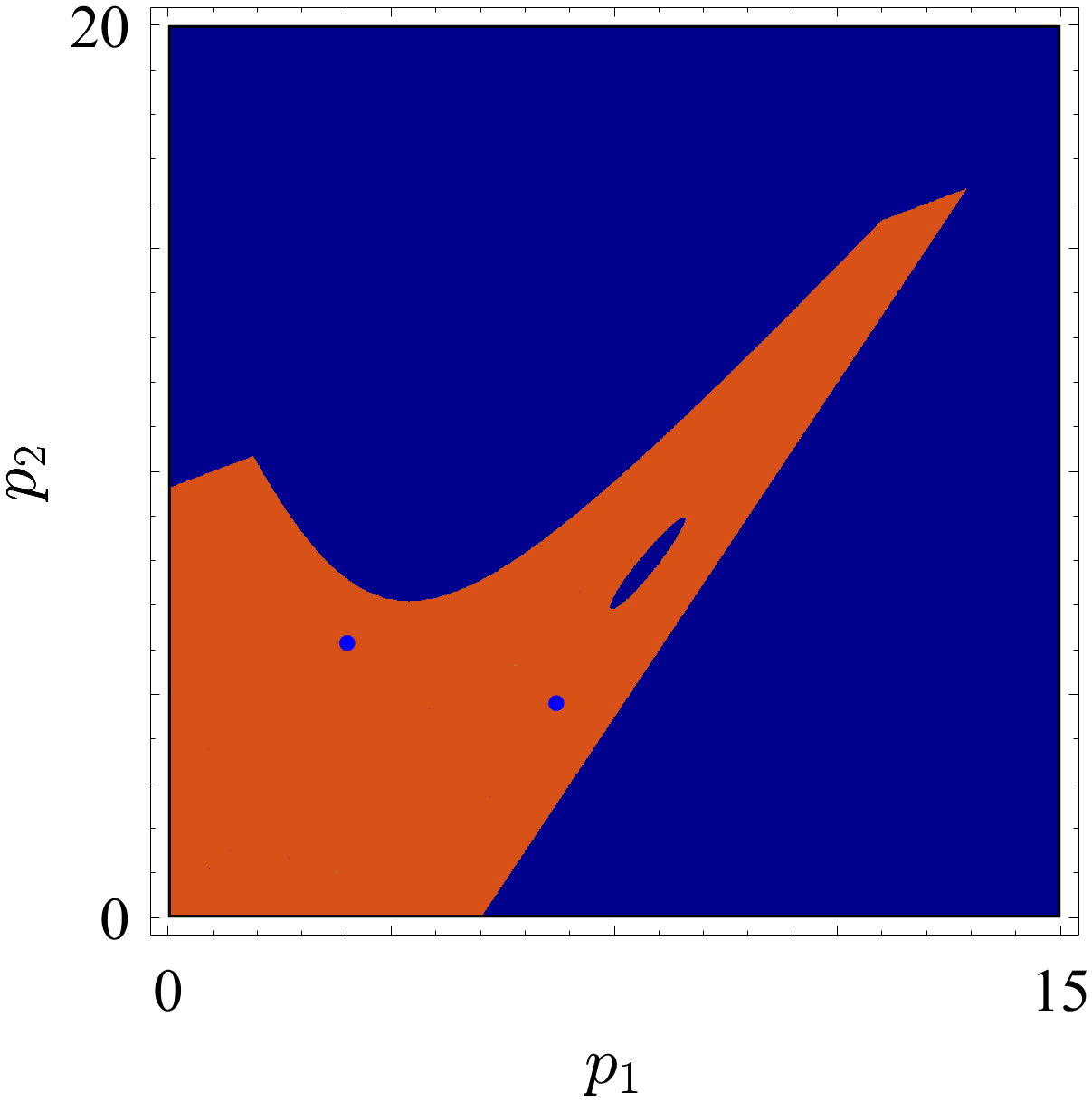}}\\
	\caption{Basins of attraction and projection of attractors in the $(p_1, p_2)$ decision space.}
	\label{fig: Fig_basin2}
\end{figure*}

As shown in Figure \ref{fig: Fig_basin1}, continuously varying the adjustment parameter $\alpha_1$ dynamically reshapes the feasible domain of the dynamic game system, as evidenced through the evolution in the basin of attraction. 
Figures \ref{fig: Fig_basina} and \ref{fig: Fig_basinb} showcase changes in the feasibility domain corresponding to stable period-one solution. Further increasing $\alpha_1$ yields more complex multistability, with Figures \ref{fig: Fig_basinc} and \ref{fig: Fig_basind} exhibiting period-doubling into feasible domains for stable period-two cycles, clearly delineated by holes when basin edges contact local bifurcation thresholds. Thereafter, Figure \ref{fig: Fig_basine} portrays period-four cycles emerging. As shown in Figure \ref{fig: Fig_basinf}, there lies an uncountably infinite set of chaotic solutions as $\alpha_1$ further increase.

Figure \ref{fig: Fig_basin1} only presents the evolution of feasible regions with varying parameter values, unveiling complex basin deformations as the decision behaviors change. Specifically, as shown in Figure \ref{fig: Fig_basin2}, mystifying basins of attraction materialize near bifurcation boundaries, encapsulating coexisting attractors and multistable domains. These multistability intersperse within different zones, which together demonstrating how minor reactive variations or perturbations to the gaming landscape can profoundly reshape the decision making behaviors.

Figure \ref{fig: Fig_basin2} shows coexisting attractors materializing as adjustment parameter change. Specifically, Figure \ref{fig: Fig_basing} portrays a stable period-two cycle. Further, an increase in the adjustment parameter elicits the concurrent emergence of both period-two and period-six cycles, delineated by distinct basins. Here, initial points in the orange domain approach the period-two attractor; while the light yellow region signifies six-cycle convergence. With the increase of adjustment parameter, the stable period-six cycle into a period-twelve cycle which is denoted by the light green area, whereas the previous period-two equilibrium persists.
Further elevating parameters and continuing observation of the dynamics, yielding a stable period-two cycle coexisting with an emergent period-nine cycle attractor, and the basin of the period-nine cycle is labeled by the light purple area. Thereafter, the period-nine cycle evolves into a chaotic attractor containing nine small orbits as the parameter increases further. Eventually, the system reverts, restabilizing at a two-cycle equilibrium. 
Surveying across feasibility spaces unveils the profound sensitivity of dynamic game systems, through subtle parameter variations that drive radical bifurcations in decision-making.

\begin{figure}[!htb]
	\centering
	\subfloat[]{\includegraphics[width = 0.65\linewidth]{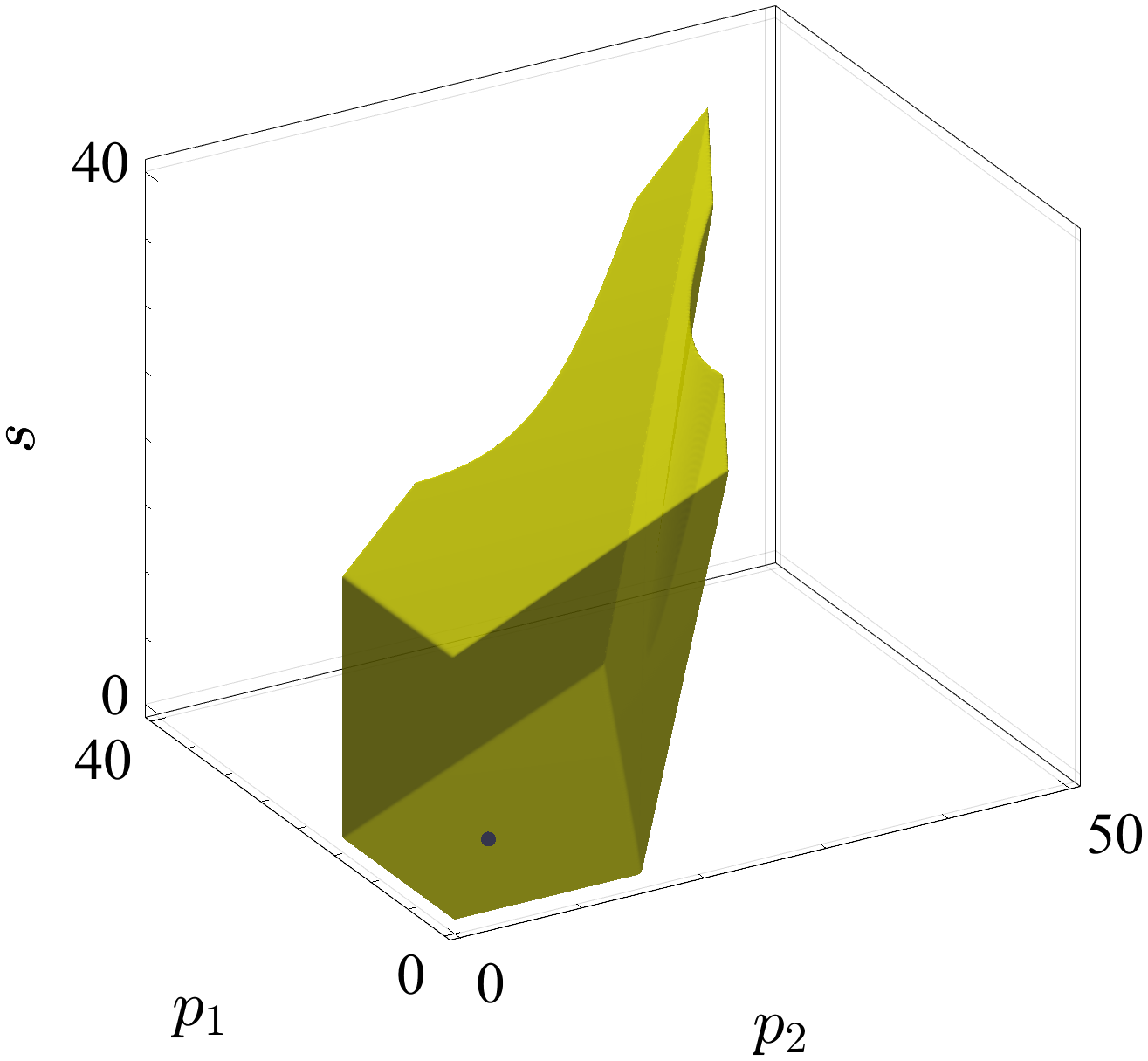}}\\
	\subfloat[]{\includegraphics[width = 0.65\linewidth]{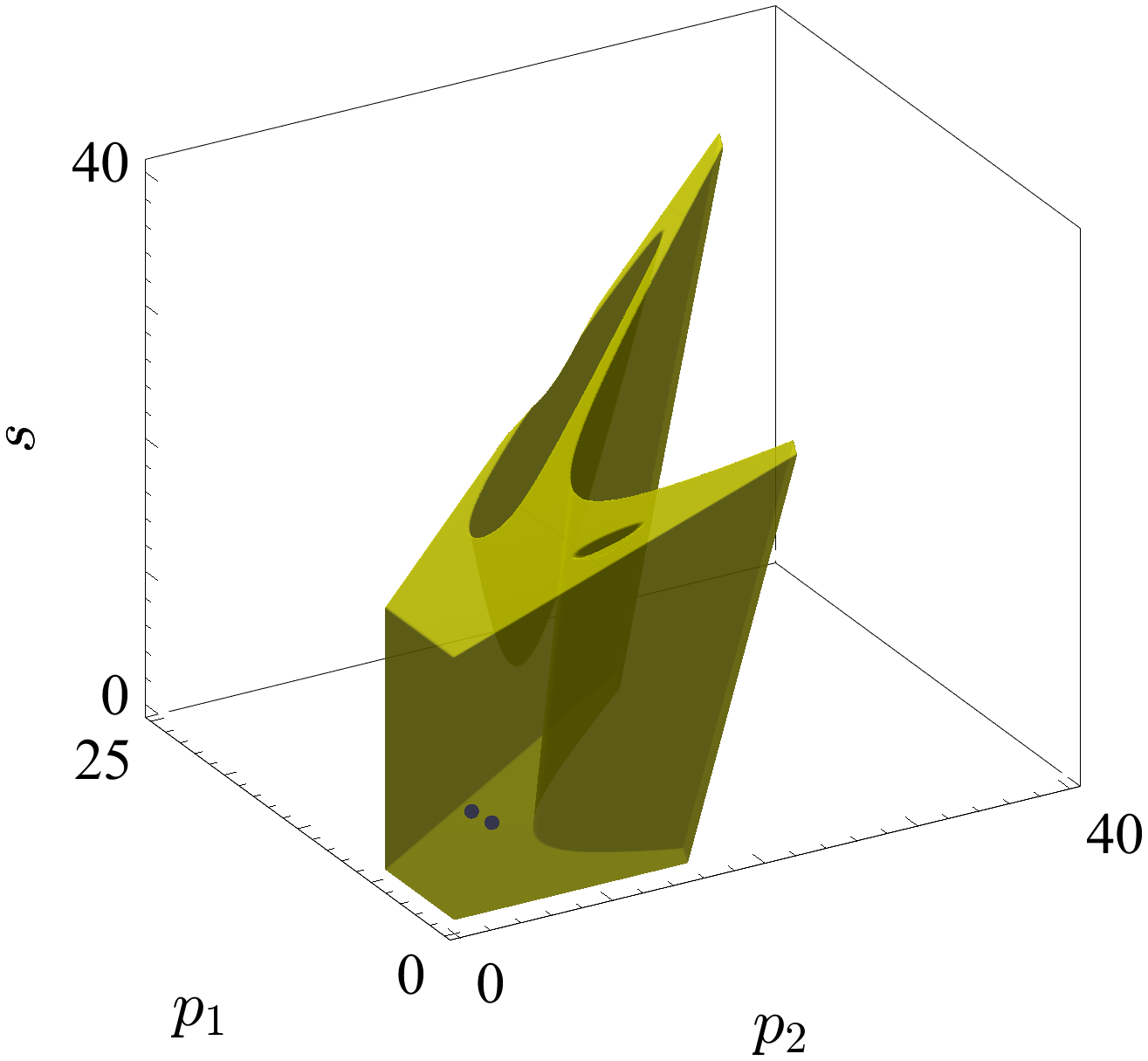}}	
	\caption{Basins of attraction and attractors in the $(p_1, p_2, s)$ decision space.}
	\label{fig: Fig_basin3d}
\end{figure}

Thus far analysis largely explored two-dimensional complex deformations in attractors and basins of attraction, including investigating occurrences of coexisting attractors. It becomes more arduous both in computation and characterizing such multifaceted landscapes when extending to higher dimensional decision spaces.  But it is helpful to understand the dynamic evolution visually. Figure \ref{fig: Fig_basin3d} proffers initial three-dimensional renderings across ($p_1, p_2, s$)-space, showcasing morphologies for both period-one cycle and period-two cycle equilibria. Though partial, these snapshots hint at extraordinary sophistication emerging even from seemingly simple game model, urging heightened methodological rigor when analyzing stability given realistic interactive contexts congruent with complexity.

As depicted in Figure \ref{fig: Fig_basin1} - \ref{fig: Fig_basin3d}, dynamic alterations across feasibility domains underscore sensitivity to initial points. This signifies that enterprises and platforms should judiciously survey potential opening conditions and perturbations that may profoundly reshape ensuing interactions. Comprehensively mapping the solution landscape assists stakeholders in maneuvering parameters to securely operate within a stable domain, thereby facilitating market stability.

\subsection{Chaos control}

\begin{figure}[!htb]
	\centering
	\subfloat[]{\includegraphics[width = 0.9\linewidth]{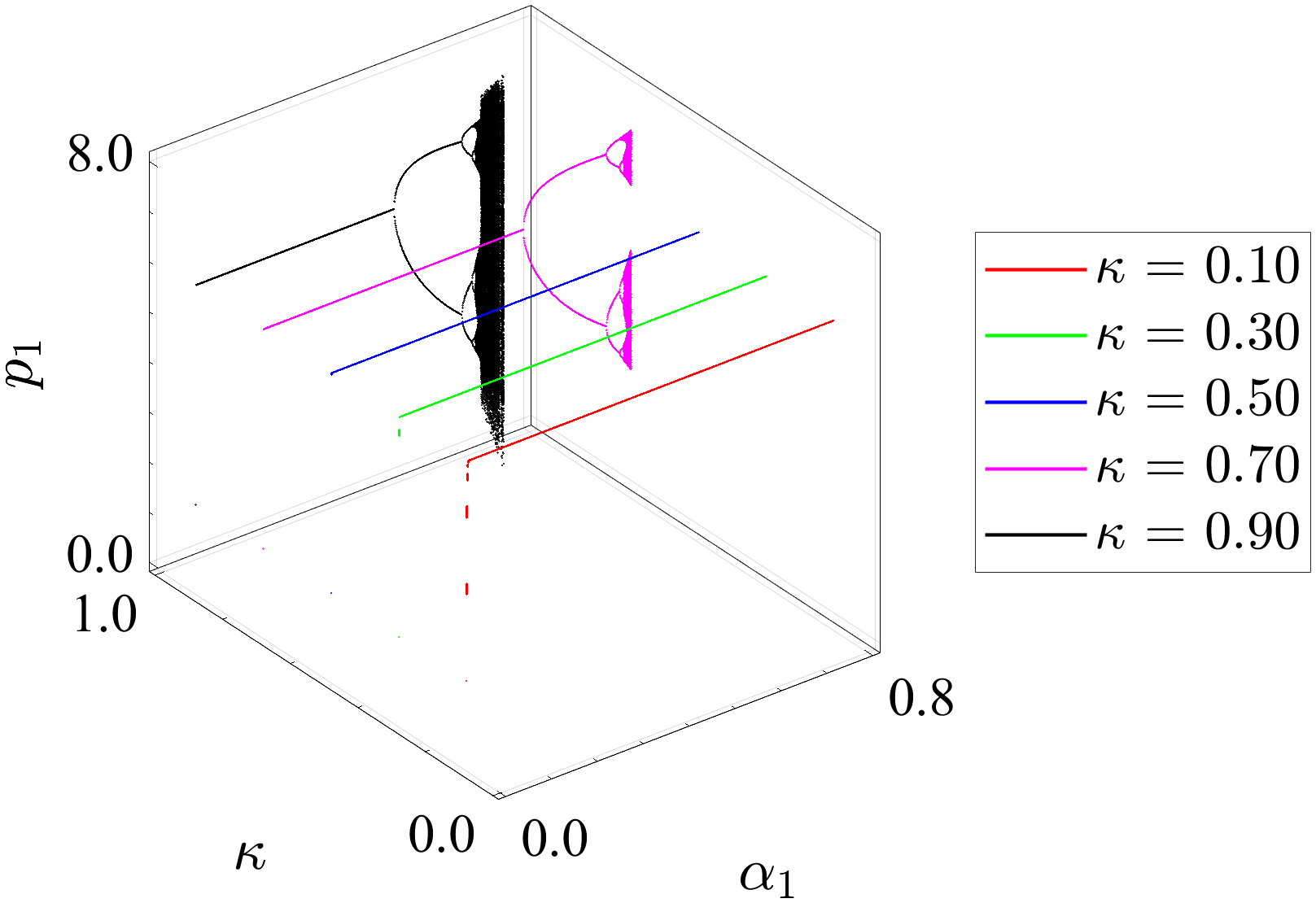}}
	\caption{Basins of attraction and attractors in the decision space.}
	\label{fig: Fig_chaos_control}
\end{figure}

Based on the study of the dynamic game system \eqref{eq05}, we know verifies minute perturbations in the initial conditions profoundly impacting long-run strategic equilibria. This sensitivity motivates exploring chaos control techniques to actively stabilize complex dynamics. Specifically, we implement a hybrid control strategy to contain complexity and avert chaos emergence\cite{Luo2003}. This adaptive mechanism functions as follows:
\begin{equation}\label{eq08}
	\begin{cases}
		p_1 (t + 1) = \kappa \left( p_1 (t) + \alpha_1 p_1 (t ) \frac{\partial \pi_1}{\partial p_1}\right) + (1 - \kappa) p_1 (t)\\
		p_2 (t + 1) = \kappa \left(p_2 (t) + \alpha_2 p_2 (t ) \frac{\partial \pi_2}{\partial p_2}\right) + (1 - \kappa) p_2 (t)\\
		s (t + 1) = \kappa \left(s (t) + \alpha_3 s (t ) \frac{\partial \pi_2}{\partial s}\right) + (1 - \kappa) s (t)\\
	\end{cases}
\end{equation}
where $\kappa \in (0, 1)$ is the control parameter. Invoking the Jury stability criteria reveals this hybrid controller successfully stabilizes system \eqref{eq05} for sufficiently small $\kappa$, as evidenced in associated bifurcation diagrams in Figure \ref{fig: Fig_chaos_control}. 

Comparisons with the uncontrolled model showcases chaos being averted under regulation when $\kappa < 0.406$. It demonstrates the controller's effectiveness in furnishing reliable equilibria to decision-makers, by steering the game system toward reliable regions even amidst unpredictability.  In other words, strategically selecting control gains enables guiding unstable dynamics toward steady, optimal equilibrium solutions. Therein, regulation not only improves strategy robustness but assists policymakers in fostering stable, efficient marketplaces amidst complex game interactions.

\section{Conclusion}\label{sec05}

This study establishes a decision-making game model for energy enterprises' data sales channels, incorporating preference information for different sales channels in consumer markets. It investigates the competitive relationship between energy enterprises and third-party platforms. In the base model, by analyzing the platform's willingness and investment intensity to extract data value, as well as the wholesale price it pays to energy enterprises, relationships between different optimal profits are obtained. It is discovered that both parties' profits are related to these key parameters. However, blindly increasing investment intensity does not necessarily improve the platform's profit, which is closely related to the wholesale cost it pays to energy enterprises. Accordingly, energy enterprises can obtain relatively higher profits by adjusting this wholesale price.

We also utilize nonlinear dynamics theory to delineate the complex evolutionary trajectories of data trading markets. The analysis reveals key factors like investment scale, wholesale pricing, and market matching degree that substantially impact revenues for both enterprises and platforms. More critically, minute parameter variations may spur abrupt transitions from stability into chaotic turbulence, urging caution in policymaking.
Thus, enterprises and platform decision-makers should judiciously formulate combinations while closely monitoring market variables, especially when introducing additional controls. Doing so assists principals in securing stable expectations and maximizing payoffs aligned with social welfare optimization. Furthermore, designing hybrid control frameworks signifies viable pathways for steering state configurations amidst unpredictability.
In summary, by quantitatively illuminating the underlying mechanisms driving multifaceted data trading ecosystems, this game theoretic analysis provides significant policy insights for principal decision-making, helping guide an increasingly digitized industry toward enhanced efficiency, sustainability, and stability.

While this study extensively analyzes the theoretical model from a game theory perspective, some limitations exist regarding practical sector data simulations, integration of additional variables, and policy implementation considerations. In the future, we will expand the model to more comprehensively study the game with all parties, including the roles of customers and government. More factors will also be considered, such as policies and market dynamics along the path of data capitalization.

\section*{Acknowledgments}
We thank the editors and anonymous reviewers for their valuable suggestions, which significantly improved the study. 

\bibliographystyle{IEEEtran}
\bibliography{bib}

\vfill

\end{document}